# Hydrodynamic Impedance Correction for Reduced-Order Modeling and Real-Time Control of Spermatozoa-Like Soft Micro-Robots for Medicine

**Ahmet Fatih Tabak[1,2*],**

[1]Mechatronics Engineering Department, Faculty of Engineering, Okan University, Istanbul, Turkey.

[2]Physical Intelligence Department, Max-Planck Institute for Intelligent Systems, Stuttgart, Germany.

**\* Correspondence:**
ahmet.tabak@okan.edu.tr; ahmetfatih.tabak@mpg-alumni.de

## Abstract

Hydrodynamic interactions play a key role in the swimming behavior and power consumption of bio-inspired and bio-mimetic micro-swimmers, Cybernetic or artificial alike. As micro-robotic devices, bio-inspired micro-swimmers require fast and reliable numerical models for robust control in order to carry out demanding therapeutic tasks as envisaged for more than sixty years. The fastest known numerical model, the resistive force theory (RFT), incorporates local viscous force coefficients with the local velocity of slender bodies in order to find the resisting hydrodynamic forces, however, omitting the induced far field altogether. In this scheme, the forces are calculated for a pure fluid resistance, however, at the expense of time-dependent hydrodynamic interaction effects. As a result, the power requirement cannot be predicted accurately although the supply of necessary power is one of the biggest obstacles impeding the micro-robotic efforts. In this study, an analysis strategy is proposed to improve the RFT-based analysis, particularly for spermatozoa and spermatozoa-inspired micro-swimmers with elastic slender tails, in order to present a practical solution to the problem. The required analytical correction and the necessary correction coefficients are based on hydrodynamic impedance analysis via intensive computational fluid dynamics (CFD) models, i.e., the time-dependent solution of 3-dimensional Navier-Stokes equations incorporated with deforming mesh and subject to conservation of mass. The CFD-based model results are then used to extract the correction coefficients with the help of a cost function written based on the hydrodynamic power required to sustain the continuous swimming of selected shape and design. The performance of the corrections embedded in the RFT-based model is finally validated against the CFD-based model by means of hydrodynamic power comparisons. It has been demonstrated that the corrected modeling approach provides rapid results with comparable accuracy for parameterized wave geometry in a predefined design space, i.e., varying wavelength and wave amplitude, presenting a fast and reliable numerical tool for real-time therapeutic applications.

## 1    Introduction

Soft robots are widely studied for various purposes as well as possible medical applications in small scales (Hines et al., 2016; Huang et al., 2016). Large body of experimental efforts have been demonstrated in the literature in the recent years on how to employ or mimic the spermatozoa in order to utilize plane wave propagation for micro-robotic actuation (Nelson et al., 2010; Peyer et al., 2013; Sitti et al., 2015; Ceylan et al., 2017); however, comprehensive theoretical models date back to the early 1950s, at which point Sir Taylor provided the analysis on the swimming sheet of infinite length conducting travelling plane wave deformations (Taylor, 1951). Effects of wave geometry are formulated with the first order perturbative approximation of the Stokes flow induced by the wave propagation. In the second paper, Taylor later expanded his analysis to predict the swimming velocity of undulating spermatozoa (Taylor, 1952). The analysis was based on the perturbative approximation of the Stokes flow generated by the undulatory motion. One year later, the novel approach of representing such a flow field by singularity distribution along a deforming slender rod is introduced by Hancock (Hancock, 1953). The same year Gray studied the internal and external force balance of undulatory motion, i.e., generation and propulsive effect of the traveling plane wave, in great detail (Gray, 1953). Later, Gray and Hancock presented the resistive force theory assuming that local deformations along a slender structure invoke local viscous resistance while the flow field is kept undisturbed in the far-field (Gray and Hancock, 1955). Eliminating all the higher-order effects and terms from the induced flow field represented by the distribution of singularities along the undulating slender body let the authors articulate the force-velocity relationship in a simple linear equation with a simple proportional constant.

The viscous flow regime assumption, i.e., a Stokes-flow solution that gave rise to the aforementioned results, omits the inertial forces, however, at the expense of accuracy of the solution. Brenner and Cox suggested corrections within the stress tensor governed by the Stokes flow around a body of an arbitrary shape to account for the incorrect force calculations along the lateral directions given by conventional analytical solutions (Brenner and Cox, 1963). It has been long known that the analytical models that make use of a variety of resistance force coefficient sets (Brennen and Winet, 1977) for simple geometries do not predict hydrodynamic forces accurately (Keller and Rubinow, 1976). Sir Lighthill articulated the slender body theory to its form as known today (Lighthill, 1976). The analysis depends on the distribution of singularities along a deforming slender body, which also gave rise to the resistive force theory (RFT). Lighthill also provided resistive force coefficients that he called the suboptimal set for micro swimmers with finite helical tails (Lighthill, 1976). Brennen and Winet provided a comparative review of resistive force theory and slender body theory with a narrative on how wave propagation is generated throughout the flagella. Authors also summarized the resistive force coefficients with respect to swimming conditions, accompanied by a detailed list of measurements on single-celled organisms conducted at the time (Brennen and Winet, 1977). Higdon took finally the slender body theory (SBT) and applied it to a representative swimmer of spherical head and a flagellum having sinusoidal plane wave propagation as means of propulsion (Higdon, 1979). Further, the author compared the results with resistive force theory coefficients in the literature by means of hydrodynamic power consumption and found that RFT calculations result in overestimated or underestimated results with respect to varying shape of the head.

Purcell put the necessity for time-irreversible action to overcome the apparent viscosity of the flow field the object is immersed in into perspective (Purcell, 1977). Furthermore, the explanation of the phenomenon is coined as the Scallop theorem. This understanding provides the motivation for research on bio-inspired swimming in micro-realm, in general. A considerable portion of the theoretical and numerical effort remained on resolving the hydrodynamic interaction of the micro-swimmer with the induced flow field. Lawrence and Weinbaum discussed that even, the oscillating or reciprocating motion of a perfect sphere is expected to cause unsteady forces due to viscous history effect and inertial effects of the flow field generated by that sphere in motion (Lawrence and Weinbaum, 1988). Lovalenti and Brady discussed that the convective effect of the viscous flow field on a submerged shape of arbitrary geometry demands a correction to unsteady Stokes solution governing the fluid-structure interaction (Lovalenti and Brady, 1995). A year later, Fauci presented a study of sperm motility based on the time-dependent solution of Navier-Stokes equations governing hydrodynamic interactions between multiple swimmers and the environment (Fauci, 1996). A solution of Navier-Stokes equations and analytical models containing inertial terms, by means of numerical methods presents a fruitful analysis opportunity as the higher order effects are no longer to be omitted for the analysis of micro-swimmers.

The hydrodynamic interaction between body and tail of a micro-swimmer was known to alter the fluid drag on the head or on the tail (Phan-Thien et al., 1987; Ramia et al., 1993). However, this indicates a possible role of near-field and far-field effects during the swimming action that were mostly dismissed for sake of asymptotical analysis. The induced flow field then studied in great detail to ascertain and quantify as to the reason for the discrepancy between the results of the Stokes-flow analysis and the more elaborate numerical solutions. Conca et al. investigated the pure added mass effect of a submerged geometry in a viscous liquid and concluded that the effect is not a direct result of viscosity or the flow field around a moving body (Conca et al., 1997). Tabak and Yesilyurt studied the flow field with consecutive shear and pressure zones generated by sinusoidal traveling plane waves in three-dimensions and reported on the effect of wave geometry on the induced flow field (Tabak and Yesilyurt, 2008). Also, squirmers, i.e., single-celled organisms equipped with cilia instead of flagella but still relying on an undulatory motion for propagation, are known to interact in the viscous flow conditions via unsteady effects (Giacché and Ishikawa, 2010). Thus the induced flow does not necessarily present a vector field of steady motion of the bulk. Wang and Ardekani showed that the Basset history force and added mass effect would be dominant against Stokes force for a simple spherical body representing a squirmer type single-celled swimmer (Wang and Ardekani, 2012). Tabak and Yesilyurt demonstrated that the RFT method needs corrections even for time-averaged velocity calculations for propulsion relying on traveling plane wave propagation (Tabak and Yesilyurt, 2013). The hydrodynamic interaction of sperm cells with nearby boundaries and the resultant effect on the swimming trajectory has been studied by means of boundary element method (BEM) based on the Stokes equation (Ishimoto and Gaffney, 2014). Felderhof and Jones analyzed the swimming of an induced flow field by a spherical body having surface distortions of wave propagation in nature concluding that the overall swimming action of such an object exhibits inertial effects of substantial contribution (Felderhof and Jones, 2018). Furthermore, authors predicted that vortex formation is expected as a part of the induced flow field in viscous fluids. Finally, Tabak articulated the time-dependent impedance correction approach for RFT-based modeling of bacteria and bacteria-inspired micro-swimmers, employing time-dependent three-dimensional Navier-Stokes equations coupled with 6-degrees-of-freedom rigid-body motion and arbitrary Lagrangian-Eulerian (ALE) mesh deformation (Tabak, 2018a). It should be also noted, as a contradictory example to the examples summed up here, Friedrich et al. reported that resistive force theory is capable of predicting the time-dependent velocity of bull sperm cells swimming close to plane boundaries (Friedrich et al., 2010), which might be attributed to the circular path as a result of the hydrodynamic interaction between the cell and the plane boundary in close proximity.

In the meantime, the wave shape for an elastic filament, i.e., the slender tail, is also investigated thoroughly over the years by means of theoretical, numerical studies, and observations. Wiggins and Goldstein articulated the elastohydrodynamic problem of plane wave propagation along elastic filaments in viscous liquids to determine the shape of deformations along the slender tail (Wiggins and Goldstein, 1998). Becker et al. studied the swimming direction of a simple reciprocating swimmer such as Purcell's three-link swimmer along with its efficiency, which found to be quite low given the viscous stresses and discussed the geometric conditions for the swimmer to follow a straight path during swimming action (Becker et al., 2003). Self-driving micro-robots relying on plane wave propagation along sheets immersed in polar liquids were also considered as viable options for propulsion (Felderhof, 2011). Koehler et al. studied the effect of plane waves of various shapes propagating along finite-length filaments without a head (Koehler et al., 2012). Authors demonstrated that the optimal swimming is based on the trade-offs between lateral degrees of freedom and the total number of waves along the filament. Previous computational efforts show that the beating of cilia, or wave propagation along a slender body, strongly depends on the geometry and interaction of the slender body with the induced flow field (Khaderi and Onck, 2012). Gadêlha studied the effect of the soft micro-swimmer on its swimming velocity by means of optimized wave deformation and varying shape of its head (Gadêlha, 2013). The author introduced a coupled mathematical model for the elastohydrodynamic and magneto-hydrodynamic problem and cast it to predict the efficient magnetic actuation parameters. Berman et al. modeled the slender tail as a series of spheres in contact and formulated the propulsive effect by means of resistive force theory and presented optimum wave geometry in terms of hydrodynamic propulsion efficiency (Berman et al., 2013). Montenegro-Johnson and Lauga numerically determined that the optimal waveform is not a sine wave but rather a complex pattern different than both sinusoidal and triangular shapes as previously thought (Montenegro-Johnson and Lauga, 2014). Furthermore, it is known that spermatozoa resort to planar wave propagation only when physical conditions are more suitable such as moving very close to a boundary, and thus, make use of helical wave propagation when it is more efficient (Woolley and Vernon, 2001; Nosrati et al., 2015). Moreover, non-sinusoidal wave patterns are also observed in nature (Woolley, 2007; Smith et al., 2009). The wave pattern and choice is affected by different physical and chemical stimuli. Computational efforts also extend to simulate the chemotaxis of such single-celled organisms to better understand the role of physical stimuli on the overall swimming behavior (Alvarez et al., 2014). Micro-swimmers are better understood to employ an intricate beating strategy than just having ordinary run-tumble phases with homogenous wave patterns. As a matter of fact, although helical and planar wave propagations are the two well-known time-irreversible actions carried out by the tails, i.e., flagella and cilia, of the single-celled organisms, there are more complex three-dimensional beating patterns accompanied with intricate resultant trajectories (Rossi et al., 2017). Furthermore, it is observed that even spermatozoa have three-dimensional trajectories due to asymmetries (Crenshaw et al., 2000).

A body of experimental studies accumulated in the literature based on spermatozoa-like micro-swimmers, in a wide spectrum of scaling, with tails varying from tens of centimeters to a few micrometers in length. Dreyfus et al. manufactured the first micro-scale cybernetic micro-swimmer towing a red-blood cell via a magneto-elastic tail made out of superparamagnetic nanoparticles and DNA strands in series (Dreyfus et al., 2005). Authors induced the plane wave propagation with the computer controlled external magnetic field. This study is an important cornerstone as it constitutes the first true biocompatible, artificial, and computer controlled micro-swimmer in the literature. Later, Roper et al. gave a very detailed analysis of the wave propagation along the magneto-elastic tail (Roper et al., 2006). Yu et al. experimented on the propulsion by elastic plane wave propagation in the cm-scale. The plane wave deformation was induced by a DC-motor driving a Scotch-yoke and lever mechanism (Yu et al., 2006). The propulsive force was measured by a strain gauge and a cantilever beam at the tip of which the cm-scale robot was anchored. In 2007, Kosa et al. proposed a swimmer with two parallel tails composed of piezoelectric laminates carrying out planar wave propagation (Kosa et al., 2007). The tail was manufactured out of three cm-scale piezo-laminates of different lengths that were calculated based on the mathematical model presented by the authors to maximize the propulsive force. Diller et al. presented programmable magnetic strips of silicon rubber embedded with magnetic particles in order to generate sinusoidal plane waves for propulsion, in mm-scale (Diller et al., 2014). Williams et al. studied a bio-hybrid micro-swimmer relying on local stresses for wave generation. In their study, authors used single or multiple cardiac cells grafted on slender PDMS filaments with geometric anisotropy and demonstrated that this approach can actually generate local deformations to induce an overall planar wave deformation which eventually results in forward-thrust (Williams et a., 2014). Khalil et al. demonstrated a spermatozoa-inspired magnetic micro-robot built via MEMS manufacturing techniques with CoNi magnetic core in its head and controlled by computer controlled oscillating weak magnetic fields (Khalil et al., 2014a). Furthermore, sperm cells are already being considered for computer controlled locomotion for cargo delivery in micro-realm (Khalil et al., 2014b). In comparison to all these studies, in 2014, Qiu et al. showed that non-reciprocal action may give net displacement if the liquid is non-Newtonian with shear thinning or shear thickening property is present (Qui et al., 2014). Medina-Sánchez et al. demonstrated the use of propulsive micro-exoskeletons for sperm cells with geometric aberrations impeding their swimming performance (Medina-Sánchez et al., 2015). Thus, sperm cells were assisted towards the target by external magnetic fields. However, this method does neither rely on plane wave deformation nor lead to effective multimodal wave propagation. Jang et al. demonstrated the swimming performance of a nano-scale filament actuated by an oscillating magnetic field (Jang et al., 2015). The filament is of anisotropic composite nature constituting single or multiple links (Ni) and hinges (polymer) in series to articulate plane wave deformations. This method allows micro-swimmers very small in length to be manufactured and yet controlled by external magnetic fields. Cheang and Kim demonstrated the manufacturing strategy of micro- and nano-robots via self-assembly of nanoparticles (Cheang and Kim, 2015). Khalil et al. manufactured and modeled spermatozoa-inspired magnetic soft-robots (Khalil et al., 2016). The micrometer-scale robots were manufactured in one step via electrospinning technique. This method is suitable for manufacturing a large number of robots of varying dimensions in one process. Xu et al. manufactured a cybernetic micro-swimmer combining drug-loaded sperm cells with an artificial magnetic extension adding computer-controlled maneuverability and guide-and-release action with tetrapod micro-structures printed within (Xu et al., 2018). Khalil et al. manufactured magnetic soft-robots with two elastic tails and demonstrated their ability to swim forward and backward by changing the actuation frequency (Khalil et al., 2018a). In the following work, Khalil et al. demonstrated the use of this feature in the independent control of multiple robots with a single actuation frequency (Khalil et al., 2018b).

MRI systems are considered as an integral part of certain micro-robotic applications in the future (Martel et al., 2009) and soft-robots for elastic tails and undulation based propulsion strategies have been envisaged as to implement MRI-controlled micro-swimmers for biomedical applications (Kósa et al., 2012). As a robotic system, one of the key issues of such a tool is the controllability which demands real-time monitoring and prediction of the trajectory. The position and velocity control of micro-robotic devices can be achieved by visual servoing (Mahoney et al., 2011; Khalil et al. 2017), ultrasound (Khalil et al., 2018c), by MRI imaging (Martel, 2017), by fluorescence imaging (Servant et al., 2015), by vision-based localization (SLAM) (Turan, 2017; Turan, 2018). However, as the size of the individual micro-swimmer gets small enough to operate near the resolution limit of the method of choice and the ability to achieve visual-servoing is undermined by the environmental complications, a model predictive observer will be required to resolve the time-dependent swimming behavior of the micro-swimmer to better control the trajectory of such a robotic device while carefully maneuvering in a network of ducts and channels filled with viscous fluids (Tabak, 2018a; 2018b).

The trajectory of an immersed soft-robot is vastly dependent on the hydrodynamics and fluid-structure interactions. The hydrodynamics of single-celled organisms and robotic micro-swimmers mimicking their swimming strategy can be represented by a linear equation simply introducing a resistance matrix for the relation between the velocity and the induced fluid force (Lauga and Powers, 2009). However, the mathematical representation of the constituents of this matrix is of the question due to the discrepancy between the Stokes-flow approximation and the higher-order effects known to surface under certain flow conditions. Yundt et al. (1975) employed resistive force coefficients to predict the trajectories of spermatozoa of Sea-urchin and rabbit, and reported that there exists considerable discrepancy between observed and calculated trajectories (Yundt et al., 1975). Moreover, kinematic and hydrodynamic modeling of spermatozoa cells and sperm-like micro-robots are dependent on the same principles as the models of bacteria and helical robots do; however, accurate kinematic models are not constructed with the same topology of equations as the motion is inherently planar if there are no geometric aberrations. Thus, although a similar CFD model governed by time-dependent Navier-Stokes subject to arbitrary-Lagrangian-Eulerian mesh configuration is employed, the hydrodynamic corrections introduced for helical swimmers earlier (Tabak, 2018a) do not hold for plane wave propagation. In this study, the previous efforts are further expanded for the planar wave propagation (Tabak and Yesilyurt, 2013; Tabak, 2018a) with time-dependent analysis: a modified method of time-dependent hydrodynamic impedance analysis based on hydrodynamic coupled with rigid-body dynamics is presented accompanied with performance comparisons against the CFD results. It is further demonstrated that the resistance matrix, modified with time-dependent hydrodynamic interaction coefficients, predicts the power consumption of such micro-swimmers with superior speed and comparable accuracy providing a fitting tool for real-time applications in comparison to slower and computationally-intensive numerical methods that are known to be suitable for time-dependent prediction of fluid forces. It is also deduced from the analysis that the swimming microrobot, in this form of modeling approach, presents a control problem fractional order in nature (Aghababa, 2014), which will be eventually beneficial for motion and gait planning of such micro-robotic swimmers (Cicconofri and DeSimone, 2016) while performing under direct computer control.

## 2    Methods

### 2.1    CFD Model

Figure 1 depicts the numerical setup, i.e., boundary conditions, for the computational fluid dynamics model accompanied with geometric variables of planar waves and frames of reference at which the fluid forces and velocities are computed. The micro-swimmer is placed in the middle of a cylindrical channel with a diameter ten times that of the head of the micro-swimmer, $R$, to decrease the hydrodynamic interaction due to the presence of the walls (Van der Sman, 2012). Length of the tail, $L$, is chosen to be two times the diameter of the head. Furthermore, the diameter of the slender tail is set to be one-tenth of the diameter of the head. As discussed earlier, it is not uncommon for micro-swimmers to exhibit complex propulsion and propagation strategies (Rossi et al., 2017; Crenshaw et al., 2000). Here, the analysis is limited to the planar wave deformation as it represents a certain propulsion method commonly shared by a certain group of artificial micro-swimmers present in the literature. Furthermore, the sinusoidal traveling plane wave is assumed to be confined to the plane on which the swimming action takes place, for sake of simplicity of the analysis and eliminating additional kinematic and hydrodynamic effects. Thus the resultant rigid-body velocity vector of the micro-swimmer has three nonzero components, i.e.,

$$\begin{bmatrix} \mathbf{V} \\ \mathbf{\Omega} \end{bmatrix} = \begin{bmatrix} V_x & V_y & 0 & 0 & 0 & \Omega_z \end{bmatrix}^{\mathrm{T}}$$

in total. Moreover, the planar wave propagates towards the +x direction while the propulsion happens to be in the −x direction in the frame of reference of the micro-swimmer. It is further assumed that, for magnetic spermatozoa-like micro-swimmers, the electromagnetic actuation system is able to follow the rigid-body rotation, $\Omega_z$, (Yu et al., 2010) while generating strong enough field to overcome the Brownian noise (Ghosh and Fischer, 2009). Under these assumptions, the equation of motion used in the CFD model is written in its own inertial frame (Spong and Vidyasagar, 1989) coupling the rigid-body dynamics with fluid dynamics over fluid-structure interaction as

$$\int_{\text{tail+head}} \begin{bmatrix} (-p\mathbf{I} + \boldsymbol{\tau})\mathbf{n} \\ (\mathbf{P} - \mathbf{x}_{com}) \times (-p\mathbf{I} + \boldsymbol{\tau})\mathbf{n} \end{bmatrix} \mathrm{d}A = \begin{bmatrix} m_{\text{tail+head}} \dfrac{\mathrm{d}\mathbf{V}}{\mathrm{d}t} \\ \mathbf{J}_{\text{tail+head}} \dfrac{\mathrm{d}\boldsymbol{\Omega}}{\mathrm{d}t} \end{bmatrix}$$

with $\mathbf{P}$ being the position of an arbitrary point along the surface of the micro-swimmer, and $\mathbf{n}$ being the surface normal at that point, at any given instant, $t$. The mass and instantaneous moment of inertia associated with the micro-swimmer are denoted by $m$ and $\mathbf{J}$. Moreover, $p$ gives the static pressure while $\boldsymbol{\tau}$ is the viscous stress tensor of an incompressible, isothermal, Newtonian liquid (Batchelor, 2005). The symbol $\mathbf{I}$ stands for 3×3 identity matrix. The instantaneous center of mass is given by the vector $\mathbf{x}_{com}$ and it is calculated for the sinusoidal deformation imposed along the elastic tail given by

$$B(x,t) = B_o\left(1 - e^{-c_L x/L}\right)\left(1 - e^{-c_{\Delta t} t/\Delta t}\right)\cos(2\pi f t - 2\pi/\lambda\, x)$$

propagating in the in the x-direction with local displacement in the y-direction as represented in Figure 1. Here, $B_o$ is the maximum wave amplitude which will develop under the influence of spatial and temporal envelope constant, $c_L$ and $c_{\Delta t}$, respectively, in order to simulate the joint between the tail and the elastic tail while eliminating the risk of warped mesh elements due to non-zero initial mesh deformation along the surface of the micro-swimmer. Here $\Delta t$ is chosen to be $t = 2.25$ and the envelope function guarantees full mesh deformation after $t = 0.25$, which will be denoted as $t_{\text{ramp}}$ hereafter. Also, the amplitude develops to its maximum within the one-tenth of the beating tail.

The wavelength is denoted by $\lambda$, and the frequency of the propagation is given by $f$. Furthermore, the total simulation time for each run is denoted by $\Delta t$. The fluidic medium between the boundaries of the micro-swimmer and boundaries of the cylindrical channel, $\Gamma(t)$, is governed by incompressible Navier-Stokes equations modified with respect to local arbitrary Lagrangian-Eulerian (ALE) mesh deformation strategy (Duarte et al., 2004) subject to conservation of mass as follows

$$\rho\left(\frac{\partial \mathbf{U}}{\partial t} + (\mathbf{U} - \mathbf{u}) \cdot \nabla \mathbf{U}\right) = -\nabla p + \mu \nabla^2 \mathbf{U}$$

$$\nabla \cdot \mathbf{U} = \mathbf{0}$$

with the convective term modified with the local mesh deformation denoted by the velocity of each mesh node, $\mathbf{u}$, in $\Gamma(t)$ including the boundaries. Hence, the mesh nodes are stationary at the channel surfaces and mesh nodes are following the resultant rigid-body motion and wave propagation superimposed wherever necessary. Furthermore, $\rho$ and $\mu$ signify the density and dynamic viscosity of the liquid, respectively, in the channel. The boundary conditions are expressed as follows

$$\mathbf{U}|_{\text{channel}} = \mathbf{u}|_{\text{channel}} = \mathbf{0}$$

$$\mathbf{U}|_{\text{head}} = \mathbf{u}|_{\text{head}} = \mathbf{V} + \boldsymbol{\Omega} \times (\mathbf{P} - \mathbf{x}_{com})$$

$$\mathbf{U}|_{\text{tail}} = \mathbf{u}|_{\text{tail}} = \mathbf{V} + \frac{\partial(\mathbf{P} - \mathbf{x}_{com})}{\partial t} + \boldsymbol{\Omega} \times (\mathbf{P} - \mathbf{x}_{com})$$

and the velocity vector for the mesh nodes between the boundaries has a linear dependency on spatial variation throughout the $\Gamma(t)$. Moreover, the channel outlets on both sides are set as open-boundary leading to zero normal force, expressed in the following form:

$$(-p\mathbf{I} + \boldsymbol{\tau})\mathbf{n}|_{\text{outlet}} = \mathbf{0}$$

The equations represented till this point are cast in the dimensionless fashion following the scaling parameters as such the head diameter, $D_{\text{body}}$, chosen as the characteristic length-scale and $1/f$ chosen as the characteristic time-scale. As a result, the scaling Reynolds number, Re, becomes

$$\text{Re} = \frac{2\pi f \rho D_{\text{body}}^2}{\mu}$$

the reciprocal of which will be used to define the dimensionless dynamic viscosity of the domain $\Gamma(t)$, which is a common practice for micro-hydrodynamic analysis of such micro-swimmers (Phan-Thien et al., 1987; Ramia et al., 1993; Tabak and Yesilyurt, 2008; Qin et al., 2011; Maniyeri et al., 2012; Tabak and Yesilyurt 2014a; 2014b; Tabak, 2018a). The Re number is defined as Re = 0.01 for the results presented in this work. Moreover, the body diameter and actuation frequency in the CFD model are selected as unity whereas the effective density also becomes unity as a result of the non-dimensionalization procedure. The CFD simulations are carried out by a commercial finite element method (FEM) package (COMSOL AB, 2015) and the domain $\Gamma(t)$ is discretized by second-order Lagrangian tetrahedral elements with a numerical degree of freedom of 300.000 per each individual simulation. The solutions are computed with PARDISO linear solver (Schenk and Gärtner, 2004) with a maximum time-step of $1/(40f)$ for two full periods after the time-ramp is finished and mesh deformation reached its maxima throughout the domain $\Gamma(t)$. It is noted that, although the numerical model takes all 6 degrees of motion into account, the resultant rigid-body motion exhibits only three degrees of freedom effectively owing to the wave geometry. As a result, the computation time required for plane wave deformations is much shorter in comparison with the helical wave propagation (Tabak, 2018a); however, simulations take an exponentially increasing amount of time with increasing wave amplitude and wavelength. Simulations were carried out for three full periods, although the analysis was carried out for the interval between the time stamps $t = t_{\text{ramp}}$ and $t = \Delta t$.

## 2.2  RFT Model

The model in question is based on the resistance matrix of micro-swimmers, and micro-objects submerged in liquids in a linear system of equations, representing a rather nonlinear phenomenon, with the help of a series of simplifications (Gray and Hancock, 1955; Happel and Brenner, 1965; Keller and Rubinow, 1976; Purcell, 1977; Tabak and Yesilyurt, 2013). The local fluid resistance, $F$, acting on a slender object of micro-dimensions is modeled with a linear relationship to its local velocity, $V$, in an arbitrary direction $i$ in the following form

$$F_i = -c_i V_i$$

and when the motion of a rigid-body is considered with all the degrees of freedom associated with it the equation is rearranged as

$$\begin{bmatrix} \mathbf{F} \\ \mathbf{T} \end{bmatrix} = -\mathbf{B} \begin{bmatrix} \mathbf{V} \\ \boldsymbol{\Omega} \end{bmatrix}$$

incorporating the possible cross-couplings between rotations and velocities along different axes based on the rigid-body kinematics. The resistance matrix $\mathbf{B}$, for a bio-inspired micro-swimmer composed of single head and single tail, can be decomposed into two 6×6 matrices that are calculated separately, i.e.,

$$\mathbf{B} = \mathbf{B}_{\text{head}} + \mathbf{B}_{\text{tail}}$$

decoupling the two objects in the analysis. This approach simply omits the near- and far-field effects of the flow field around the object, unless the resistance matrix coefficients are modeled in a way to account for such phenomena. However, the common practice in the literature is to introduce the effect due to the presence of solid boundaries for blunt objects or flagella moving in close proximity (Happel and Brenner, 1965; Higdon and Muldowney, 1995; Lauga et al., 2006) only. The resistance matrix for the spherical head shown in Figure 1 can be expressed with the help of three 3×3 matrices in the following form

$$\mathbf{B}_{\text{head}} = \begin{bmatrix} \mathbf{D}_{\text{T}} & -\mathbf{D}_{\text{T}}\mathbf{S} \\ \mathbf{S}\mathbf{D}_{\text{T}} & \mathbf{D}_{\text{R}} \end{bmatrix}$$

taking the kinematic couplings into account. Here, $\mathbf{D}_{\text{T}}$ is the diagonal matrix of translational resistance coefficients, $\mathbf{D}_{\text{R}}$ is the diagonal matrix of rotational resistance coefficients, and $\mathbf{S}$ denoting the cross-product operation written with respect to the center of mass of the micro-swimmer. In this study, the center of mass of the micro-swimmer and the volumetric center of the spherical head are assumed to coincide, effectively rendering all elements of matrix $\mathbf{S}$ to be zero for the head. This assumption is also in good agreement with heads of homogeneous magnetic properties being actuated by external magnetic fields (Khalil et al., 2016). Under the previous assumptions, the translational and rotational matrices for the sphere are written as

$$\mathbf{D}_{\text{T}} = \begin{bmatrix} -6\pi\mu R & 0 & 0 \\ 0 & -6\pi\mu R & 0 \\ 0 & 0 & 0 \end{bmatrix}$$

$$\mathbf{D}_{\text{R}} = \begin{bmatrix} 0 & 0 & 0 \\ 0 & 0 & 0 \\ 0 & 0 & -8\pi\mu R^3 \end{bmatrix}$$

with the proper drag coefficients (Berg, 1993), respectively. On the other hand, the total 6×6 resistance matrix of the beating tail is found by a more elaborate way of integrating the local forces over the entire length as

$$\mathbf{B}_{\text{tail}} = \int_L \begin{bmatrix} \mathbf{R}\mathbf{C}\mathbf{R}^{\text{T}} & -\mathbf{R}\mathbf{C}\mathbf{R}^{\text{T}}\mathbf{S} \\ \mathbf{S}\mathbf{R}\mathbf{C}\mathbf{R}^{\text{T}} & -\mathbf{S}\mathbf{R}\mathbf{C}\mathbf{R}^{\text{T}}\mathbf{S} \end{bmatrix} \mathrm{d}x$$

Here, the 3×3 matrix $\mathbf{R}$ denotes the time and position dependent local rotation matrix from Frenet-Serret coordinates throughout the beating tail to the coordinate frame residing at the center of mass of the micro-swimmer, i.e., $\mathbf{R} = [\mathbf{t}\ \mathbf{n}\ \mathbf{b}]$ where $\mathbf{t}$ is the tangential, $\mathbf{n}$ is the normal, and $\mathbf{b}$ is the binormal direction unit-vectors on an arbitrary location along the tail as given in Figure 1 (Hanson and Ma, 1995; Roper et al., 2006). The $\mathbf{S}$ matrix, here, also stands for the cross-product operation but for any arbitrary location along the beating tail with respect to the center of mass of the micro-swimmer. Finally, the 3×3 matrix $\mathbf{C}$ holds the local resistance coefficients for the beating tail. There are several coefficient sets in the literature (Gray and Hancock, 1955; Lighthill, 1975; Keller and Rubinow, 1976; Lighthill, 1976; Brennen and Winet, 1977; Johnson and Brokaw, 1979), and here in this study the following set is used for the analysis:

$$\mathbf{C} = \begin{bmatrix} \dfrac{-2\pi\mu}{\ln\left(\dfrac{2\lambda}{r}\right) - 2.9} & 0 & 0 \\ 0 & \dfrac{-4\pi\mu}{\ln\left(\dfrac{2\lambda}{r}\right) - 1.9} & 0 \\ 0 & 0 & 0 \end{bmatrix}$$

is chosen for the RFT model. The matrix has no resistance coefficient for the bi-normal direction given that there will be no contribution of the binormal component as it solely points the z-axis. The simplifications on the resistance matrices are also validated by the results of two-dimensional swimming trajectory obtained by the previous CFD model given strict geometric symmetry on the plane of wave propagation (see Figure 2).

However, the reduced order mathematical model articulated thus far considers the velocity kinematics and omits the hydrodynamic interactions between the head and the tail. The flow field induced by the overall motion of the deforming tail introduces phase information to the linear relationship for the head. Also, there exists the hydrodynamic interaction along the beating tail itself introducing the phase information to the local force calculations. This phase information is found to be a function of time for helical wave propagation (Tabak, 2018a) and a very similar analysis will be carried out here to ascertain the time-dependent characteristics of this phenomenon for planar wave propagation.

Although the same steps of analysis will be carried out, in comparison with the previous findings, the modified $\mathbf{B}_{\text{head}}$ and $\mathbf{C}$ matrices are expected to represent an interaction somewhat different than what has been articulated before owing to the fact that the entire motion is confined to the xy-plane as depicted in Figure 2. The trajectory of bacteria and bacteria-like swimmers, on the other hand, are expected to be helical by default (Tabak, 2018a). The time-dependent force-velocity relation can be expressed in terms of phasors, similar to that of the electric or acoustic impedance (Kino, 1987; Hambley, 2002), in order to account for the hydrodynamic impedance associated with any two velocity components of a different kind, i.e., $V_i$ and $\Omega_j$. After substantiating the phase and amplitude difference between the CFD-based and conventional-RFT-based force curves with the process of visual inspection (see Figure 3), the subsequently modified matrices are chosen to be of the following form

$$\mathbf{B}_{\text{head}}^{*} = \begin{bmatrix} -6\pi\mu R\Upsilon_{\text{b},x,\text{T}} & 0 & 0 & 0 & 0 & 0 \\ 0 & -6\pi\mu R\Upsilon_{\text{b},y,\text{T}}\cos(\phi_{y,\text{T}}) & 0 & 0 & 0 & 6\pi\mu R\Upsilon_{\text{b},y,\text{T}}\sin(\phi_{y,\text{T}}) \\ 0 & 0 & 0 & 0 & 0 & 0 \\ 0 & 0 & 0 & 0 & 0 & 0 \\ 0 & -8\pi\mu R^3\Upsilon_{\text{b},z,\text{R}}\sin(\phi_{z,\text{R}}) & 0 & 0 & 0 & -8\pi\mu R^3\Upsilon_{\text{b},z,\text{R}}\cos(\phi_{z,\text{R}}) \end{bmatrix}$$

and

$$\mathbf{C}^* = \begin{bmatrix} \dfrac{-2\pi\mu}{\ln\left(\dfrac{2\lambda}{r}\right) - 2.9}\cos(\phi_{\text{tail}}(t)) & \dfrac{4\pi\mu}{\ln\left(\dfrac{2\lambda}{r}\right) - 1.9}\sin(\phi_{\text{tail}}(t)) & 0 \\ \dfrac{-2\pi\mu}{\ln\left(\dfrac{2\lambda}{r}\right) - 2.9}\sin(\phi_{\text{tail}}(t)) & \dfrac{-4\pi\mu}{\ln\left(\dfrac{2\lambda}{r}\right) - 1.9}\cos(\phi_{\text{tail}}(t)) & 0 \\ 0 & 0 & 0 \end{bmatrix}$$

$$\mathbf{B}^*_{\text{tail}} = \int_0^{\alpha L} \begin{bmatrix} \Upsilon_{\text{tail,T}}\mathbf{R}\mathbf{C}^*\mathbf{R}^{\mathrm{T}} & -\Upsilon_{\text{tail,T}}\mathbf{R}\mathbf{C}^*\mathbf{R}^{\mathrm{T}}\mathbf{S} \\ \Upsilon_{\text{tail,R}}\mathbf{S}\mathbf{R}\mathbf{C}^*\mathbf{R}^{\mathrm{T}} & -\Upsilon_{\text{tail,R}}\mathbf{S}\mathbf{R}\mathbf{C}^*\mathbf{R}^{\mathrm{T}}\mathbf{S} \end{bmatrix} \mathrm{d}L$$

$$\Upsilon_{\text{tail,T}} = \begin{bmatrix} \Upsilon_{\text{t},x,\text{T}} & 0 & 0 \\ 0 & \Upsilon_{\text{t},y,\text{T}} & 0 \\ 0 & 0 & 0 \end{bmatrix}$$

$$\Upsilon_{\text{tail,R}} = \begin{bmatrix} 0 & 0 & 0 \\ 0 & 0 & 0 \\ 0 & 0 & \Upsilon_{\text{t},z,\text{R}} \end{bmatrix}$$

with the amplitude-correction terms, $\Upsilon$, and the phase-correction terms, $\phi$, incorporated in the model following different approaches. It is important to note that, this scheme is not the only possible solution but one of the likely recipes in order to seek a suitably modified version of the resistance matrices $\mathbf{B}^*_{\text{head}}$ and $\mathbf{B}^*_{\text{tail}}$. It is also important to acknowledge that these corrections are expected to have different values, and possibly different formulae, with the presence of a boundary in close proximity.

## 3  Results

### 3.1  Post Processing

Theoretically, all of the amplitude- and phase-correction coefficients mentioned here can be hypothesized as an oscillatory function of time, i.e., attaining different values at each time-step in one full period and repeating the same pattern for each consecutive period. For sake of simplicity, this will be done so for the phase-correction coefficients of the beating tail, only; the rest will be given in the time-averaged fashion. The choice shall be based upon the error margins of the application.

The performance of the amplitude and phase-corrections introduced in the resistance matrices of the head and the tail of the micro-swimmer are studied against the hydrodynamic power consumption results. The correction factors in question are introduced to these matrices under the swimming conditions discussed in this study and are found with the help of hydrodynamic power calculations for CFD model, $\Pi_{\text{CFD}}$, for conventional resistance matrix approach, $\Pi_{\text{RFT}}$, and for the modified resistance matrix, $\Pi^*_{\text{RFT}}$, respectively. These computations are accompanied with the minimization of the respective cost functions, all cast in time-dependent fashion individually. Hence the following equations to predict the power consumption are solved

$$\Pi_{\text{CFD,tail}}(t) = \Pi_{\text{CFD,tail},x}(t) + \Pi_{\text{CFD,tail},y}(t) + \Pi_{\text{CFD,tail},z}(t)$$

for time-dependent phase-correction search on the tail, where each element on the right-hand side that are given as

$$\Pi_{\text{CFD,tail,x}}(t) = \text{abs}\left( \int_{\text{tail}} \left( (-p\mathbf{I} + \boldsymbol{\tau})\mathbf{n} \cdot \begin{bmatrix} V_x \\ 0 \\ 0 \end{bmatrix} \right) \mathrm{d}A \right)$$

$$\Pi_{\text{CFD,tail,y}}(t) = \text{abs}\left( \int_{\text{tail}} \left( (-p\mathbf{I} + \boldsymbol{\tau})\mathbf{n} \cdot \left( \begin{bmatrix} 0 \\ V_y \\ 0 \end{bmatrix} + \frac{\partial(\mathbf{P} - \mathbf{x}_{com})}{\partial t} \right) \right) \mathrm{d}A \right)$$

$$\Pi_{\text{CFD,tail,z}}(t) = \text{abs}\left( \int_{\text{tail}} \left( ((\mathbf{P} - \mathbf{x}_{com}) \times (-p\mathbf{I} + \boldsymbol{\tau})\mathbf{n}) \cdot \begin{bmatrix} 0 \\ 0 \\ \Omega_z \end{bmatrix} \right) \mathrm{d}A \right)$$

used for time-averaged amplitude-correction on the tail. These equations are followed by similar definitions given as

$$\Pi_{\text{CFD,head,x}}(t) = \text{abs}\left( \int_{\text{head}} \left( (-p\mathbf{I} + \boldsymbol{\tau})\mathbf{n} \cdot \begin{bmatrix} V_x \\ 0 \\ 0 \end{bmatrix} \right) \mathrm{d}A \right)$$

$$\Pi_{\text{CFD,head,y}}(t) = \text{abs}\left( \int_{\text{head}} \left( (-p\mathbf{I} + \boldsymbol{\tau})\mathbf{n} \cdot \begin{bmatrix} 0 \\ V_y \\ 0 \end{bmatrix} \right) \mathrm{d}A \right)$$

$$\Pi_{\text{CFD,head,z}}(t) = \text{abs}\left( \int_{\text{head}} \left( ((\mathbf{P} - \mathbf{x}_{com}) \times (-p\mathbf{I} + \boldsymbol{\tau})\mathbf{n}) \cdot \begin{bmatrix} 0 \\ 0 \\ \Omega_z \end{bmatrix} \right) \mathrm{d}A \right)$$

for the head of the micro-swimmer. The RFT-based equivalents of these power relations can also simply be given as

$$\Pi^*_{\text{RFT,tail}}(\phi, t) = \Pi^*_{\text{RFT,tail,x}}(\phi, t) + \Pi^*_{\text{RFT,tail,y}}(\phi, t) + \Pi^*_{\text{RFT,tail,z}}(\phi, t)$$

$$\Pi^*_{\text{RFT,tail,x}}(\phi, t) = \text{abs}\left( \mathbf{B}^*_{\text{tail}}(\phi, t) \begin{bmatrix} \mathbf{V} + \frac{\partial(\mathbf{P} - \mathbf{x}_{com})}{\partial t} \\ \boldsymbol{\Omega} \end{bmatrix} \cdot \begin{bmatrix} V_x \\ 0 \\ 0 \\ \mathbf{0} \end{bmatrix} \right)$$

$$\Pi^*_{\text{RFT,tail,y}}(\phi, t) = \text{abs}\left( \mathbf{B}^*_{\text{tail}}(\phi, t) \begin{bmatrix} \mathbf{V} + \frac{\partial(\mathbf{P} - \mathbf{x}_{com})}{\partial t} \\ \boldsymbol{\Omega} \end{bmatrix} \cdot \begin{bmatrix} \begin{bmatrix} 0 \\ V_y \\ 0 \end{bmatrix} + \frac{\partial(\mathbf{P} - \mathbf{x}_{com})}{\partial t} \\ \mathbf{0} \end{bmatrix} \right)$$

$$\Pi^*_{\text{RFT,tail,z}}(\phi, t) = \text{abs}\left( \mathbf{B}^*_{\text{tail}}(\phi, t) \begin{bmatrix} \mathbf{V} + \frac{\partial(\mathbf{P} - \mathbf{x}_{com})}{\partial t} \\ \boldsymbol{\Omega} \end{bmatrix} \cdot \begin{bmatrix} \mathbf{0} \\ 0 \\ 0 \\ \Omega_z \end{bmatrix} \right)$$

$$\Pi^*_{\text{RFT,head},x}(\phi, t) = \text{abs}\left( \mathbf{B}^*_{\text{tail}}(\phi, t) \begin{bmatrix} \mathbf{V} \\ \mathbf{\Omega} \end{bmatrix} \cdot \begin{bmatrix} V_x \\ 0 \\ 0 \\ \mathbf{0} \end{bmatrix} \right)$$

$$\Pi^*_{\text{RFT,head},y}(\phi, t) = \text{abs}\left( \mathbf{B}^*_{\text{tail}}(\phi, t) \begin{bmatrix} \mathbf{V} \\ \mathbf{\Omega} \end{bmatrix} \cdot \begin{bmatrix} 0 \\ V_y \\ 0 \\ \mathbf{0} \end{bmatrix} \right)$$

$$\Pi^*_{\text{RFT,head},z}(\phi, t) = \text{abs}\left( \mathbf{B}^*_{\text{tail}}(\phi, t) \begin{bmatrix} \mathbf{V} \\ \mathbf{\Omega} \end{bmatrix} \cdot \begin{bmatrix} \mathbf{0} \\ 0 \\ 0 \\ \Omega_z \end{bmatrix} \right)$$

in all directions in the frame of reference of the micro-swimmer, with $\mathbf{0} = [0\ 0\ 0]^{\text{T}}$. The power consumption with conventional RFT approach is also written similarly, however, with unmodified resistance matrices, i.e., $\mathbf{B}_{\text{head}}(t)$ and $\mathbf{B}_{\text{tail}}(t)$. The respective cost functions are given in the form of the least square error approach

$$I_{\text{tail}}(\phi, t) = \sum_{t=t_{\text{ramp}}}^{\Delta t} \left( \Pi_{\text{CFD,tail}}(t) - \Pi^*_{\text{RFT,tail}}(\phi, t) \right)^2$$

for time-dependent phase-correction search on the tail in symbolic form for each time-step of each CFD simulation, and in the form of ratios of respective time-averaged hydrodynamic power

$$I_{\text{tail},\{x,y,z\}}(\phi, \Upsilon) = \sum_{t=t_{\text{ramp}}}^{\Delta t} \Pi_{\text{CFD,tail},\{x,y,z\}}(t) \bigg/ \sum_{t=t_{\text{ramp}}}^{\Delta t} \Pi^*_{\text{RFT,tail},\{x,y,z\}}(\phi, t)$$

$$I_{\text{head},\{x,y,z\}}(\phi, \Upsilon) = \sum_{t=t_{\text{ramp}}}^{\Delta t} \Pi_{\text{CFD,head},\{x,y,z\}}(t) \bigg/ \sum_{t=t_{\text{ramp}}}^{\Delta t} \Pi^*_{\text{RFT,head},\{x,y,z\}}(\phi, t)$$

in order to extract the correction factors in the time-averaged fashion for the rest. Also, here it must be acknowledged that each direction indices, i.e., x, y, and z, corresponds to one mode of motion, i.e., translation for the first two and rotation for the third, respectively, due to the unique nature of combination of actuation and rigid-body motion of the micro-swimmer in question. Moreover, in these final post-processing analyses, i.e., solution of the said cost function equations, should be in the following order: First, the amplitude-correction should be set as unity at all time-steps, i.e., $\Upsilon(t) = 1$, thus the phase-correction terms, $\phi(t)$ and $\phi$, should be determined. The reverse order would result in a recursive search as each force component is a periodic function with its own phase information different than its constituent functions, therefore the resultant amplitude is not a simple product of the global maxima of the constituent functions. Furthermore, the cross-couplings between the degrees of freedom result in superimposing hydrodynamic effects from two different axes further shifting the position of the global extrema to the final periodic functions that are hydrodynamic force and power. Once the said phase-corrections are properly obtained, they should be substituted into the same cost functions and the necessary amplitude-corrections can now simply be resolved with the help of respective power terms, in time-dependent or time-averaged fashion. Furthermore, it is but a concern of accuracy whether one chooses to implement the said cost functions all in the form time-averaged or time-dependent fashion or not as will be discussed at length in the Discussion section.

Furthermore, it is important to acknowledge that it might not be feasible to carry out the procedure in the reverse order, i.e., first obtaining the amplitude-corrections, then obtaining the phase-corrections. Such course of action will indeed require additional steps to obtain the actual amplitude-correction, $\Upsilon$, given that superimposing the velocity components, which are in the form of sinusoidal functions such as depicted in Figure 2, with an additional phase information will yield an arbitrary amplitude for the resultant force components. Figure 4 further presents the complete algorithm to the numerical and symbolic procedure followed in this study as explained till here.

Finally, the time-averaged hydrodynamic power consumption is calculated via integration over time as shown

$$\langle \Pi^*_{\{\text{CFD,RFT}\},\{\text{tail,head}\},\{x,y,z\}}(\phi) \rangle = f \int_{1/f} \Pi^*_{\{\text{CFD,RFT}\},\{\text{tail,head}\},\{x,y,z\}}(\phi, t)\, \mathrm{d}t$$

followed by normalization over the entire period, which is a nondimensionalized unit-time in the simulations.

### 3.2 Corrections and Power Comparisons

The following results present the correction coefficients accompanied by respective performance comparisons against CFD results and conventional RFT-models in terms of predictions of hydrodynamic power consumption. Wave amplitude is parameterized such that $B_o = \{0.025\ 0.05\ 0.075\ 0.1\ 0.125\ 0.15\ 0.175\ 0.2\ 0.225\ 0.25\ 0.275\ 0.3\ 0.325\ 0.35\ 0.375\ 0.4\ 0.425\}$ with $L/\lambda$ being fixed at 2, and the wavelength is parameterized such that $L/\lambda = \{0.75\ 1\ 1.25\ 1.5\ 1.75\ 2\ 2.25\ 2.5\ 2.75\ 3\ 3.25\ 3.5\ 3.75\ 4\ 4.25\}$ with $B_o$ being fixed at 0.2. Thus, a total of 32 separate CFD simulations are employed to obtain the time-dependent force and time-dependent velocity components which are collated in the form of text files. The analysis and the search for the correction coefficients are carried out on in MATLAB environment. The data is directly uploaded from the text files and is then cast in the procedure explained in the previous sections.

Figure 5 presents the time-dependent phase-correction coefficients, $\phi(t)$, for the beating tail. The phase-correction exhibits a much stable behavior when compared with the helical wave propagation (Tabak, 2018a), in general. The phase-correction is found to fluctuate with a small amplitude and an average below zero for all wavelengths, $\lambda$, and for all wave amplitudes, $B_o$, studied within the selected design space. Figure 5(A) depicts that once the wavelength shortens beyond $L/\lambda = 2.25$, the phase-correction, $\phi(t)$, becomes much smaller on the negative side dictating that the induced flow field does not have an effect as significant as when the length of the wave becomes comparable to that of the length of the tail. Figure 5(B) depicts that, with smaller wave amplitudes the phase-correction becomes obsolete. However, as the wave amplitude increases, the phase-correction becomes more useful. Furthermore, it is observed that when the wave amplitude, $B_o$, diminishes or becomes comparable to the radius of the head, $R$, the phase-correction coefficient, $\phi(t)$, does not exhibit oscillations in time. It is noted that the time is scaled by $1/f$, in simulations. Thus, one complete period takes 1 unit-time in CFD simulations.

Figure 6 presents the time-averaged amplitude-correction for the force component along the x-direction, i.e., the direction of swimming, $\Upsilon_{\text{x-translation}}$, acting on the beating tail. Results for varying wavelength and wave amplitude are normalized by their respective absolute maximum as given in associated legends, thus clearly showing the variation. It is important to note that the amplitude correction for both cases is on the order of $10^{-5}$. This is a significant indicator showing that the selected local resistance coefficients are overestimating the forces greatly when the phase-correction coefficients are in place. Furthermore, the majority of the data points do correspond to correction coefficients of an even smaller order of magnitude. It is also important to notice that the fluctuation around zero, as in both Figure 6(A) and Figure 6(B), indicates that the average values might not fully represent this system; however, it will be evident that these results will suffice for the given design space, for all intents and purposes.

Figure 7 presents the amplitude correction for force components acting on the beating tail in the lateral directions, i.e., lift and rotation along the xy-plane, denoted by $\Upsilon_{yz\text{-}\{\text{translation, rotation}\}}$. Figure 7(A) and Figure 7(C) demonstrates fluctuations mostly around zero, once again hinting at the oscillating nature of the fluid-structure interaction throughout the beating tail with respect to varying wavelength. On the other hand, Figure 7(B) depicts the monotonic relationship of the correction coefficients with respect to varying wavelength with $L/\lambda = 2$. It is important to note that, different values and profiles can be expected as the ratio $L/\lambda$ attains different values. Figure 7(D) also expresses an intricate relationship of the amplitude-correction with local minima and maxima: It is observed that there are distinct regions, again, hinting that the induced flow field coupled with the rigid-body dynamics shows different characteristics suggesting intricate hydrodynamic interactions between the flow field generated by the moving body and the flow field generated by the beating tail attached to it. Once more, the coefficients shown in all subplots are normalized with their respective absolute maximum as presented in associated legends. Akin to the previous figure, the order of magnitude of the correction factors dictates that the selected resistance coefficients overestimate the local forces when the phase corrections are cast.

Figure 8 represents the performance of the correction coefficients, which are hypothesized for the beating tail alone, against the CFD and conventional RFT results in terms of time-averaged hydrodynamic power consumption, $< \Pi >$, with the help of parameterized wavelength. Hence, the corrections for the head are not involved in these results. The conventional RFT analysis using the resistance matrices without the correction coefficients greatly overestimates the hydrodynamic power required to sustain the wave propagation as depicted in Figure 8(A) and in Figure 8(B). On the other hand, the improved RFT model shows great potential by predicting the power requirement in the same order as the CFD model provides. However, the performance of the corrections slightly deteriorates at the boundaries of the design space, i.e., with very small and very large wavelengths (see Figure 8(B)); however, the corrections yield an almost one-to-one match in several cases. It is further important to notice that, the power consumption is predicted with high precision at $L/\lambda = 2$, which reflects on the results presented in Figure 9 positively.

Figure 9 presents the performance of the correction coefficients, which are hypothesized for the beating tail alone, against the CFD and conventional RFT results in terms of time-averaged hydrodynamic power consumption, $< \Pi >$, with the help of parameterized wave amplitude. Hence once again, the corrections for the head are not included in these results. The performance of the improved RFT analysis employing the correction coefficients are with an overwhelming advantage over the conventional RFT analysis without the correction coefficients as depicted in Figure 9(A) and in Figure 9(B): The conventional RFT analysis predicts the hydrodynamic power requirement with a staggering error of 7 order of magnitude. Moreover, the variation exhibited by CFD analysis and by the improved RFT analysis shows the linear relationship of hydrodynamic power requirement to wave amplitude variations as opposed to the variations in wavelength as given by Figure 8(B).

Figure 10 presents the time-averaged phase-correction, $\phi$, for the spherical head being pushed forward by the beating tail. These corrections are specified for the lateral directions alone. For lateral translation, the phase correction is close to $2\pi$, or $0\pi$, in all cases except $L/\lambda = \{1.5, 3.5, 3.75\}$ with $B_o$ being fixed at 0.2. However, it can be seen that there lies a monotonic decrease in phase-correction with decreasing wavelength and with decreasing wave amplitude, accompanied by small oscillations on both accounts. Presence of half waves renders the conventional RFT analysis accurate in terms of phase-lag between time-dependent CFD-based results and RFT-based results for lateral translations with respect to varying wavelength, as illustrated by Figure 10(A). On the other hand, the hydrodynamic power due to the lateral torque exhibits a slightly stronger dependence on the phase-correction coefficients, as presented in Figure 10(C). The phase-correction for varying wave amplitude is found to be constant for lateral translation and lateral rotation, although the former happens to predict $\phi$ near $2\pi$, i.e., ~$1.75\pi$, as the former predict $\phi$ near $0\pi$, i.e., ~$0.25\pi$ as presented by Figure 10(B) and Figure 10(D).

Figure 11 presents the amplitude correction, $\Upsilon_x$, for the force component acting on the head along the x-direction, i.e., the direction of swimming. The amplitude correction for both varying wavelength and varying wave amplitude seems to be fixed around $\Upsilon_{x,translation} = 0.007$ with insignificant fluctuations, except for $L/\lambda = 0.75$ and $B_o/R = 0.05$. This also indicates that wave geometry is not directly responsible for the correction within the studied design space, which suggests that the tail length, $L$, and the tail radius, $r$, could be directly responsible as long as the wave amplitude, $B_o$, is not larger than the radius of the head, $R$.

Figure 12 presents the amplitude correction for force components acting on the head in the lateral directions, i.e., lift and rotation along the xy-plane. Each subplot is normalized with its respective associated absolute maximum. Figure 12(A) shows that the amplitude correction for lateral translation, $\Upsilon_{yz-translation}$, has a period of oscillation with a period of 1 full wavelength for $L/\lambda = [1\ 4]$ and for each said period, there lies a local maxima at the presence of half waves, i.e., $L/\lambda = \{1.5, 2.5, 3.5\}$. The plot suggests that this behavior might continue beyond the limits of the design space of choice. A similar behavior can be seen in Figure 12(C) for the corrections to the lateral rotation, $\Upsilon_{yz-rotation}$: Here, it is not easy to distinguish local maxima with the seemingly-random fluctuations. However, it is apparent that the correction factor tends to increase with decreasing wavelength. It is important to note that the correction for lateral rotation is found to be three orders of magnitude smaller than the correction for lateral translation, for the spherical head of the micro-swimmer. This indicates that presence of the beating tail amplifies the drag in lateral directions, as opposed to the drag-force in the forward direction (see Figure 11) while attenuating the resistance to rigid-body rotation. Furthermore, Figure 12(B) suggests that the lateral drag slowly decreases with increasing wave amplitude and it drops well below half of an isolated sphere which indicates that the flow field invoked by the beating tail contributes to the lateral push of the head. On the other hand, it is demonstrated by Figure 12(D) that the drag to the rigid-body rotation is greatly decreased, i.e., on the order of 5, yet increases one order of magnitude with increasing wave amplitude within the design space studied in this study.

Figure 13 presents the performance of the correction coefficients, which are hypothesized for the head alone, against the CFD and conventional RFT results in terms of time-averaged hydrodynamic power consumption, $< \Pi >$, with the help of parameterized wavelength. Akin to the previous performance comparisons, the correction coefficients predicted for the beating tail are not used to calculate these results. The conventional RFT analysis predicts an optimum region within the design space of choice where the hydrodynamic power consumed by the head of the micro-swimmer can be minimized, as presented in Figure 13(A). On the other hand, Figure 13(B) dictates that the power requirement constantly but slowly increases with the decreasing wavelength. Furthermore, the improved RFT analysis employing the correction coefficients predicts the hydrodynamic power with enhanced accuracy but with slight error due to an apparent fluctuation with respect to varying wavelength. Furthermore, it is noted that the conventional RFT method overestimates the power requirement with an error on the order of 2.

Finally, Figure 14 presents the performance of the correction coefficients, which are hypothesized for the head alone, against the CFD and conventional RFT results in terms of time-averaged hydrodynamic power consumption, $< \Pi >$, with the help of parameterized wave amplitude. Once again, it is acknowledged that the corrections obtained for the beating tail are not employed in the following results to single out and differ the importance of each respective correction. According to the results given by Figure 14(A) and Figure 14(B), in general, all three approaches predict an almost constant increase in power consumption with a similar profile, however, with a different order. Similar to the previous plots, the improved RFT analysis provides results in good agreement with the results of the CFD simulations as given in Figure 14(B). Moreover, the increasing power consumption to sustain the wave propagation with increasing wave amplitude is expected considering the increasing lateral drag, and constant forward drag which dictates increasing forward velocity in return.

Table 1 provides the root-mean-square (RMS) errors for conventional and improved RFT results, against the CFD-based results. Results are presented in terms of minimum, maximum, average, and standard deviation calculated with respect to the cost functions discussed in the previous section, for individual respective parameterized wave dimension. The improvement attained for the beating tail is on the orders of O(6)-O(8) whereas the improvement is on the orders of O(3)-O(4) for the spherical body. The overwhelming difference observed in the results of the beating tail demonstrates how the amplitude of time-dependent force and power calculations deviate greatly when the phase-correction is not incorporated considering that the amplitude corrections are smaller than unity as presented in Figure 6 and Figure 7.

## 4    Discussion

It may seem mathematically efficient to eliminate higher order effects; however, this may cause a complete overestimation of power requirements in time-domain, and subsequently, misrepresentation of the hydrodynamic forces acting on a spermatozoon-inspired micro-swimmer. It is our intention to transmit the necessary power to control the swimming speed and velocity while maintaining the control of the motion and position of the micro-robotic tool in the therapeutic operations of the future. This study elaborates on a mathematical and numerical scheme on how to predict the power consumption of such a micro-robotic tool under constraints of the real-time applications. The time-dependent solution of full Navier-Stokes equations is superior only if the computational power is apt to resolve the spatial and temporal behavior. Such solutions are expensive in terms of equipment and time. However, the finite-element method and its derivatives are not suitable for obtaining quick results. The fastest method available is the resistive force theory (RFT) which can predict the associated forces using a simple linear relationship, however, with significant error. It is important to acknowledge that, the conventional RFT analysis might be successful in some swimming conditions; however, one must check the viability of the solution by means of computational fluid dynamics or molecular dynamics simulations whenever possible. The RFT method will predict the time-averaged behavior and the maxima with greater accuracy when it comes to kinematic analysis. It has been shown in this work that the improved RFT method using the correction coefficients inspired by the impedance analogy will be successful at dynamic analysis too.

The conventional resistance matrix does not incorporate the out-of-phase behavior of the surface velocity and hydrodynamic force components associated with the overall motion of the boundaries on the swimming robot. Thus, it predicts in-phase periodic functions for force and velocity. Furthermore, the total power along the beating tail is found via superimposing contributions from two or more directions due to the cross-couplings between the main axis as manifested in the overall resistance matrix. Thus, the conventional RFT method falls short in predictions of the location of the peaks, i.e., the timestamps when the extrema occur, of hydrodynamic force and power consumption. In contrast to this, the associated phase-corrections improve the resistance matrix to the point that this issue is properly resolved and the extrema of said periodic functions are correctly predicted in time-domain.

The particular correction coefficients and the modified resistance matrix, such as it is, are not the only possible improvement to the RFT approach. It is important to demonstrate that the beating tail is interacting with the induced flow field in such a way that there exists a phase-lag between local force and local velocity components along its surface. The time-averaged coefficients might be suitable in some designs and time-dependent coefficients in the other; however, one needs to study the impact of these corrections on the accuracy of the force and power requirements. Larger wave amplitudes were particularly tough to handle as to the numerical convergence and warped mesh issues. Regardless, one important assumption used in this work is the homogeneity of the wave shape throughout the beating tail, except near the joint. Hence, the effect of structural damping on the local deformations is omitted. It is also important to acknowledge that the study presented here does not take the effect of nearby boundaries into account. Depending on the shape of the confining boundaries, the corrections might appear in different form and value. The flow field induced by the beating tail and the rigid-body motion of the micro-swimmer has a symmetry-plane only when neutral buoyancy and isolated swimming conditions are satisfied. Once that symmetry is broken, the induced flow field is expected to exhibit a different behavior that demands a dedicated study to resolve the interactions and associated impedance alongside.

The analysis scheme presented here makes use of a fast analytical method and improves it to the point that it can be cast in model-based control and used as a model-based observer in real-time applications. The correction factors introduce the fractional-order nature of the micro-swimmer which will also be made use of when implementing suitable control algorithms. The one important drawback is the necessity to carry out the analysis for each and every design of interest to determine the correction coefficients offline, which will take significant computational effort. However, the power consumption and the associated forces can be predicted online quickly and accurately via this RFT-based approach whereas the conventional approach will fall short on predicting time-dependent dynamics of such systems.

## 5    Conflict of Interest

*The author declares that the research was conducted in the absence of any commercial or financial relationships that could be construed as a potential conflict of interest.*

## 6    Author Contributions

AFT designed and conducted the computational fluid dynamics simulations, formulated the mathematical model for the study, wrote the code necessary to carry out the analysis laid out in this manuscript, and drafted the manuscript.

## 7 Funding

## 8 Acknowledgments

## 10    Figures

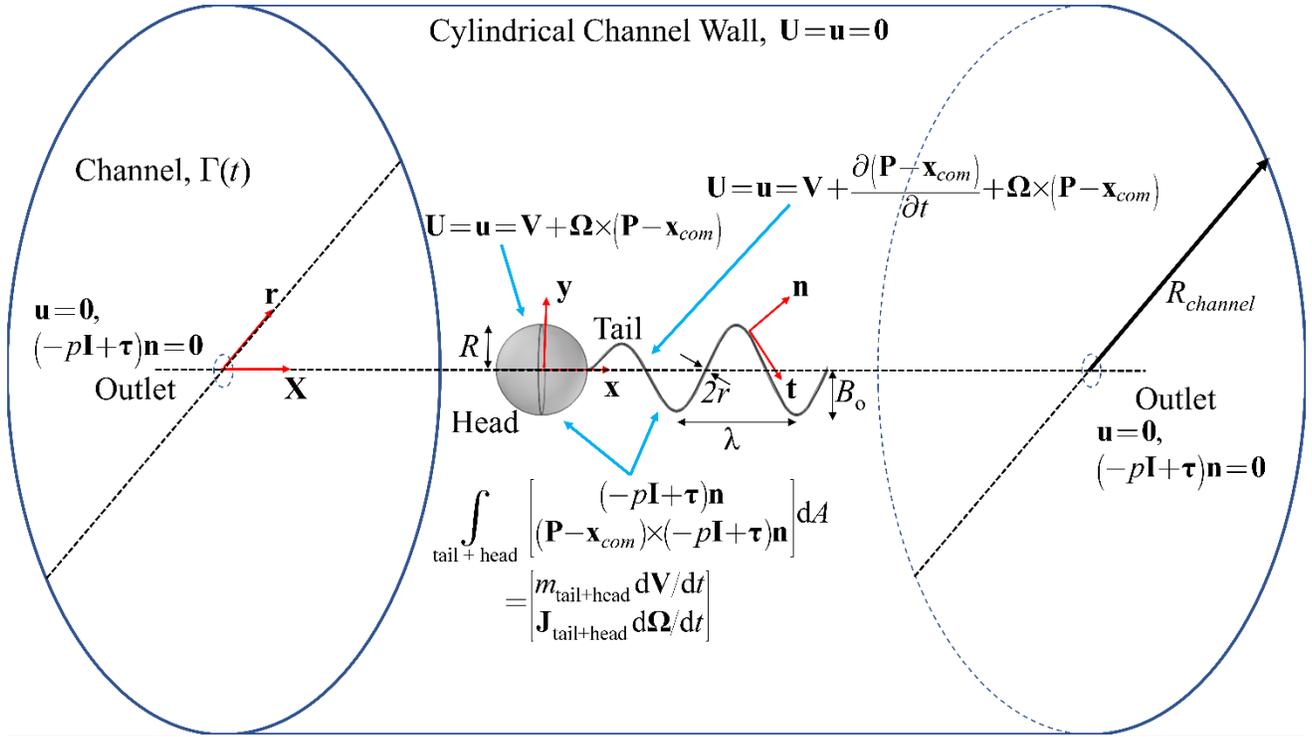

**Figure 1:** Numerical set-up of the CFD model: Boundary conditions on the spermatozoa-like micro-swimmer and the cylindrical channel walls that mark the liquid medium, $\Gamma(t)$. The fluid forces are coupled with rigid-body dynamics.

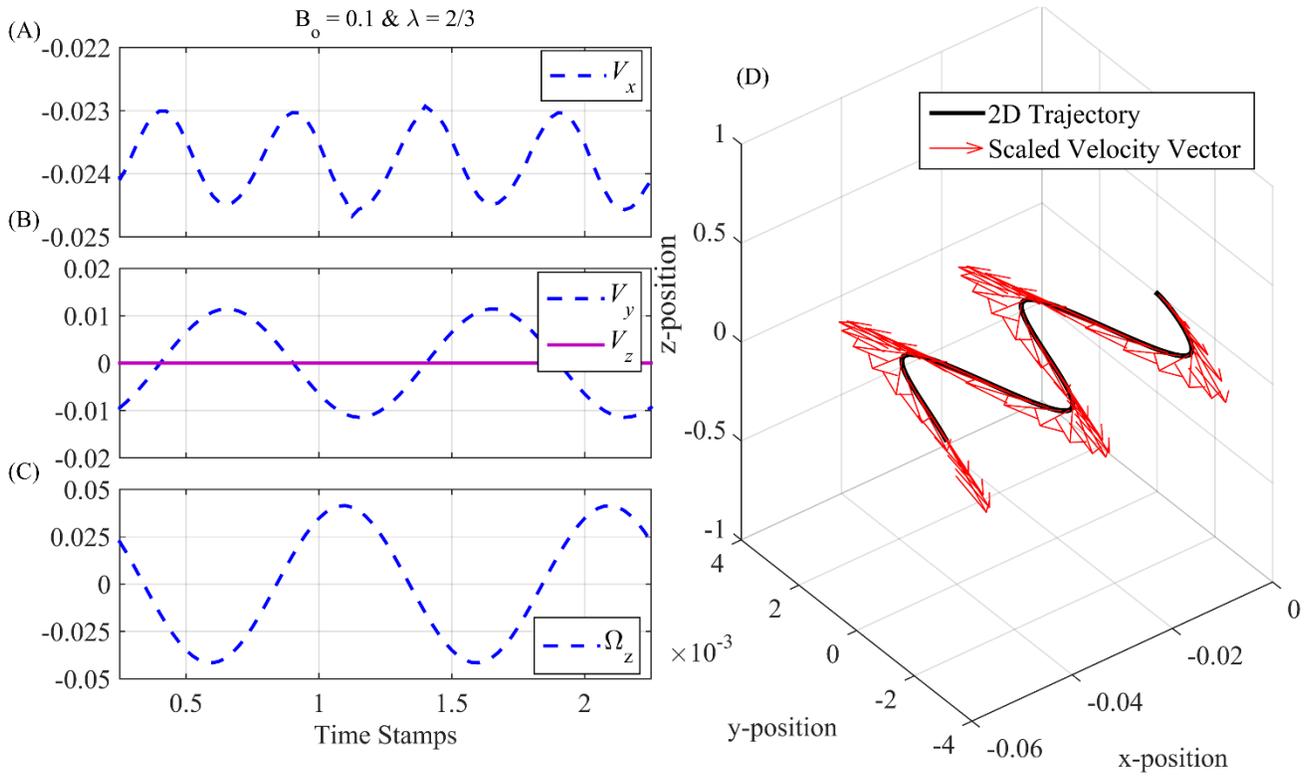

**Figure 2:** Time-dependent velocity and trajectory predictions for the micro-swimmer for a representative case of $L/\lambda = 3$ and $B_\mathrm{o} = 0.1$. The CFD simulation is governed by time-dependent Navier-Stokes incorporated with ALE mesh deformation and subject to conservation of mass.

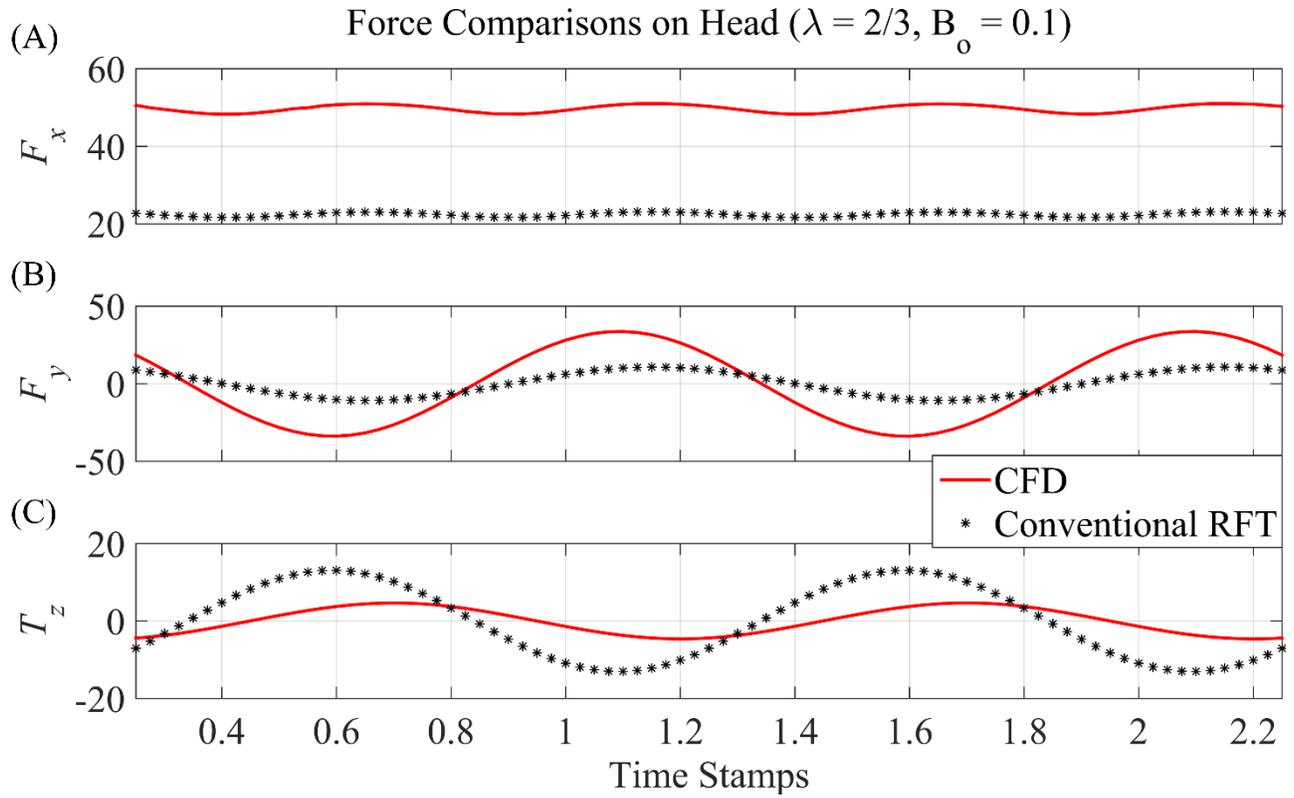

**Figure 3:** Time-dependent force calculations for the spherical head of the micro-swimmer for a representative case of $L/\lambda = 3$ and $B_o = 0.1$. The RFT calculations are carried out without any corrections and compared against the results of time-dependent CFD simulation.

**Hydrodynamic Impedance Correction for Reduced-Order Modeling and Real-Time Control of Spermatozoa-Like Soft Micro-Robots for Medicine**

**Algorithm:** Determination of the Correction Factors

1: **do** %% main loop for numeric & symbolic analysis
2:   **do** %% Start the CFD & rigid-body dynamics analysis for the micro-swimmer
3:     **if** $t = 0$ **then**
4:       Initialize: $\mathbf{V}(t=0) = \mathbf{U}(t=0) = \mathbf{u}(t=0) = \mathbf{0}$ & $\omega(t=0) = 2\pi f$.
5:     **else**
6:       $\omega = 2\pi f$.
7:       Compute the total hydrodynamic force induced with the beating tail and by the swimmer's rigid body velocity vector, $\mathbf{V}$.
8:       Compute the rigid-body acceleration of the micro-swimmer.
9:       Integrate the acceleration vector to find the current velocity and position vectors.
10:     **end**
11:     Update surface velocity boundary conditions on the beating tail with computed swimming velocities, i.e., $\mathbf{U} = \mathbf{V} + d\mathbf{P}/dt + \Omega \times \mathbf{P}$, for $t = t + \delta t$.
12:     Update the surrounding mesh with ALE, for $t = t + \delta t$.
13:     Update the velocity on the moving/deforming surfaces of the swimmer.
14:   **while** $0 \le t \le t_{\text{final}}$
15:   Extract the velocity and force components with respect to simulation time array.
16:   Write the total resistance matrix of the head in symbolic form with correction coefficients.
17:   Write the total resistance matrix of the tail in symbolic form with correction coefficients.
18:   Write the local hydrodynamic power consumption equation for the tail in symbolic form.
19:   Write the hydrodynamic power consumption equation for the head in symbolic form.
20:   Construct empty array for phase-correction coefficient, $\phi_{\text{tail}}(t)$.
21:   **do** %% Start the symbolic RFT analysis for time-dependent phase-correction
22:     Set all the amplitude-corrections $\Upsilon(t) = 1$ and substitute them in all symbolic equations.
23:     Substitute the rigid-body velocity components $\mathbf{V}(t)$ and $\mathbf{U}(t)$ in all symbolic equations.
24:     Substitute current time $t$ in all symbolic equations.
25:     Perform the symbolic integration of the local hydrodynamic power consumption equation of the tail over its length.
26:     Write the cost functions $I_{\text{tail}}(\phi, t)$.
27:     Search for the phase-correction, $\phi_{\text{tail}}(t)$, minimizing the cost function $I_{\text{tail}}(t)$.
28:   **while** $0.25 \le t \le \Delta t$
29:   **do** %% Start the symbolic RFT analysis for time-averaged amplitude-correction
30:     Substitute the current phase-correction, $\phi_{\text{tail}}(t)$, in the original symbolic power consumption equation of the tail.
31:     Substitute the rigid-body velocity components $\mathbf{V}(t)$ and $\mathbf{U}(t)$ in all symbolic equations.
32:     Substitute current time $t$ in all symbolic equations.
33:     Perform the symbolic integration of the local hydrodynamic power consumption equation of the tail over its length.
34:     Sum up the power requirement for the head of the micro-swimmer.
35:   **while** $0.25 \le t \le \Delta t$
36:   Write the cost functions $I_{\text{tail},\{x,y,z\}}(\phi, \Upsilon)$ and $I_{\text{heat},\{y,z\}}(\phi)$.
37:   Search for the amplitude-corrections, $\Upsilon_{\text{tail},\{x,y,z\}}$, minimizing the cost function $I_{\text{tail},\{x,y,z\}}(\phi, \Upsilon)$.
38:   Search for the phase-corrections, $\phi_{\text{head},\{y,z\}}$, minimizing the cost functions, $I_{\text{heat},\{y,z\}}(\phi)$, separately.
39:   Substitute $\phi_{\text{head},\{y,z\}}(t)$ in all the previous symbolic power consumption equations.
40:   Write the cost functions $I_{\text{heat},x}(\Upsilon)$ and $I_{\text{heat},\{y,z\}}(\phi, \Upsilon)$.
41:   Search for the amplitude-corrections $\Upsilon_{\text{head},\{x,y,z\}}$, minimizing the cost functions, $I_{\text{heat},x}(\Upsilon)$, $I_{\text{heat},\{y,z\}}(\phi, \Upsilon)$, separately.
42: **while** $1 \le \text{count} \le \text{total number of simulations}$

**Figure 4:** The concatenated algorithm to the entire procedure as laid out throughout within this study, with subsequent steps of the CFD and RFT analyses.

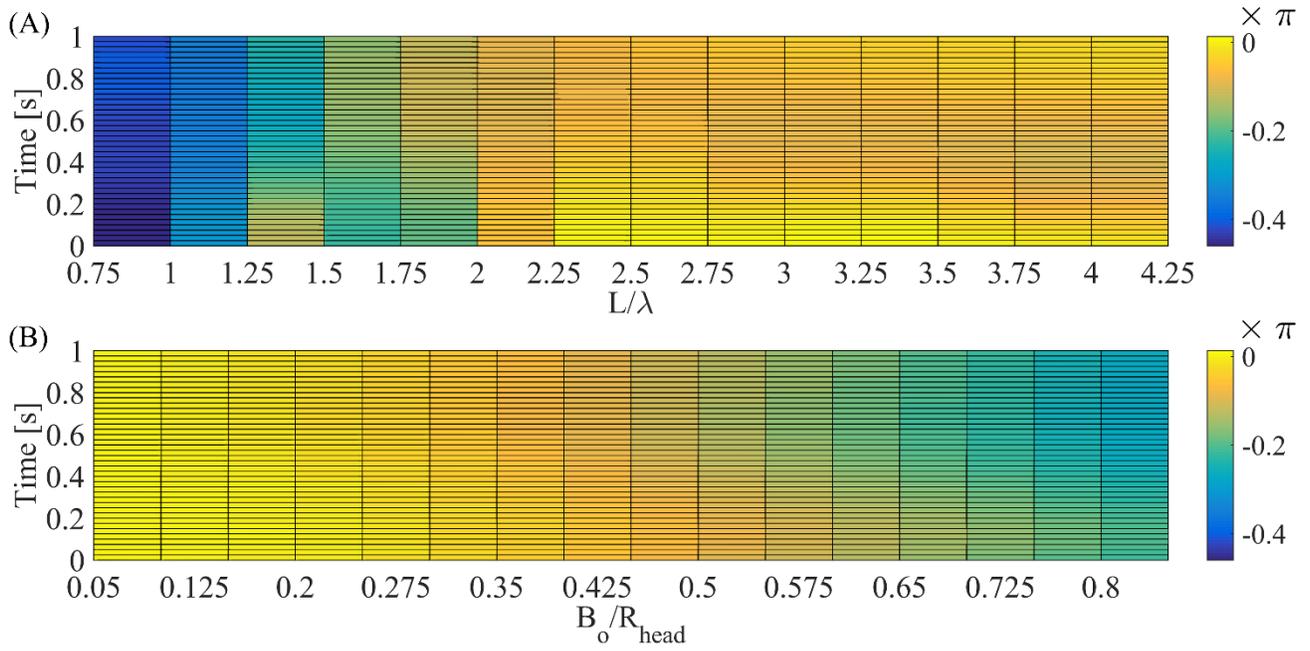

**Figure 5:** Time-dependent phase-correction coefficients for the resistance matrix of the beating tail: (**A**) the dependence on the wavelength; (**B**) dependence on the wave amplitude. Axes are presented in the dimensionless form.

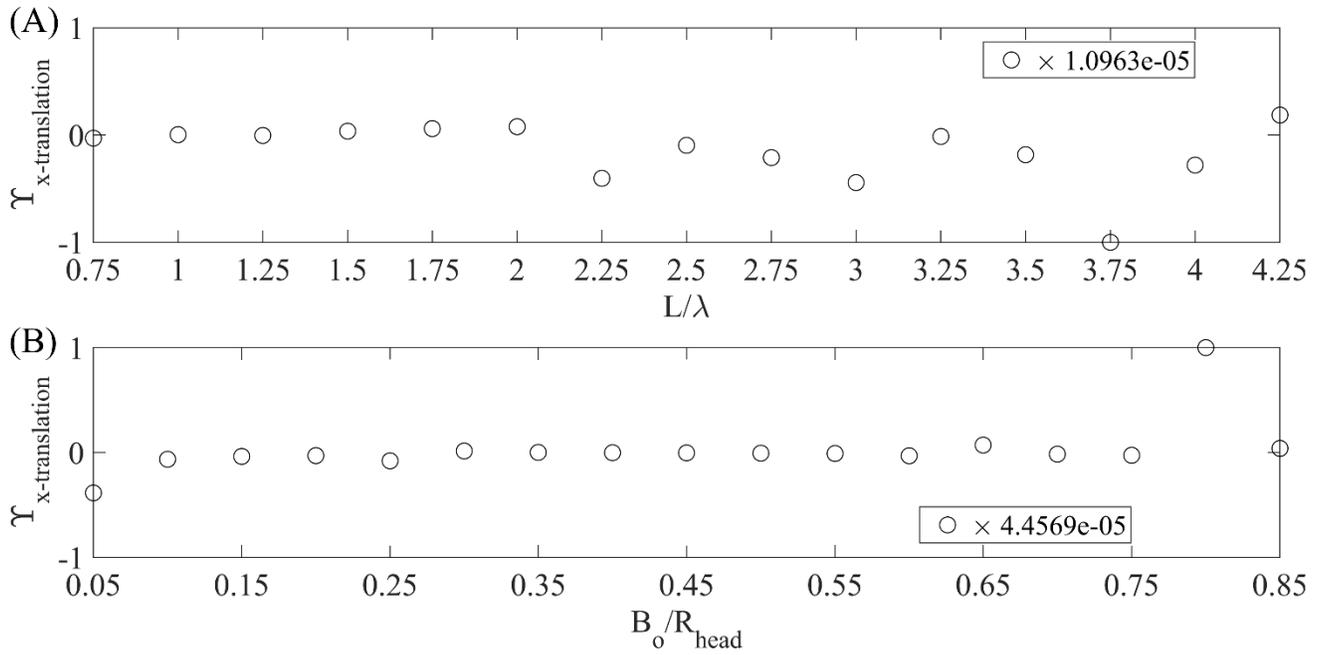

**Figure 6:** Time-averaged amplitude-correction coefficients for resistance matrix of the beating tail; the correction for the force components along the direction of swimming: (**A**) the dependence on the wavelength; (**B**) dependence on the wave amplitude. Axes are presented in the dimensionless form.

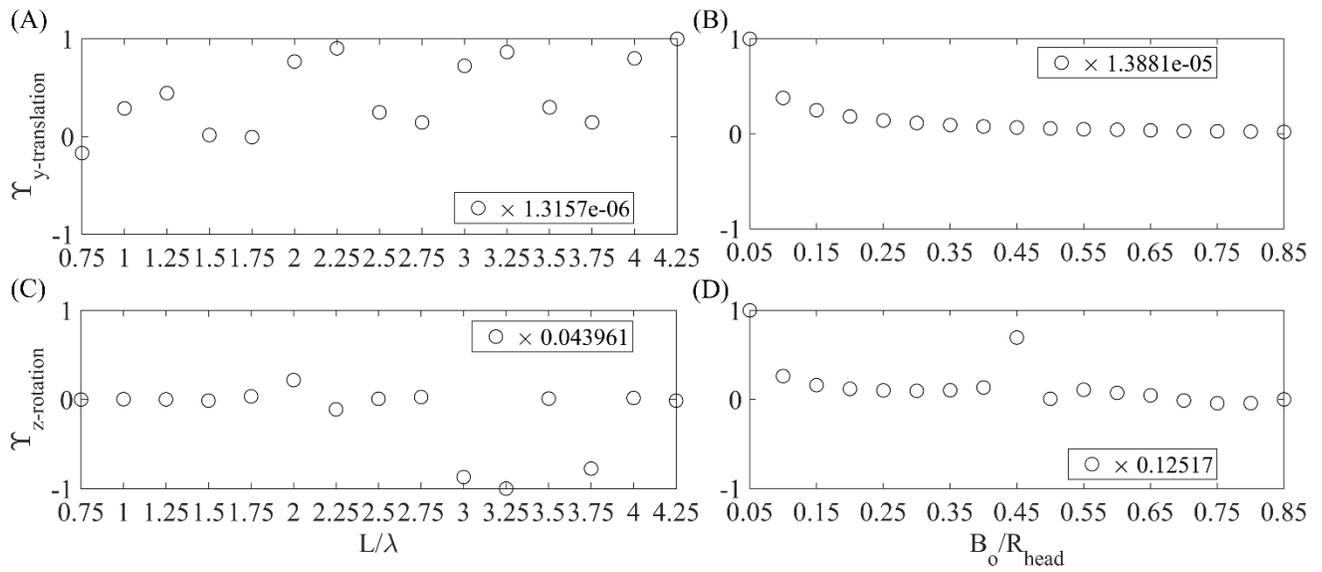

**Figure 7:** Time-averaged amplitude-correction coefficients for the resistance matrix of the beating tail: (**A**) the correction to the force associated with the lateral translation with respect to varying wavelength; (**B**) the correction to the force associated with the lateral translation with respect to varying wave amplitude; (**C**) the correction to the torque associated with the lateral rotation with respect to varying wavelength; (**D**) the correction to the torque associated with the lateral rotation with respect to varying wave amplitude. Axes are presented in the dimensionless form.

**Hydrodynamic Impedance Correction for Reduced-Order Modeling and Real-Time Control of Spermatozoa-Like Soft Micro-Robots for Medicine**

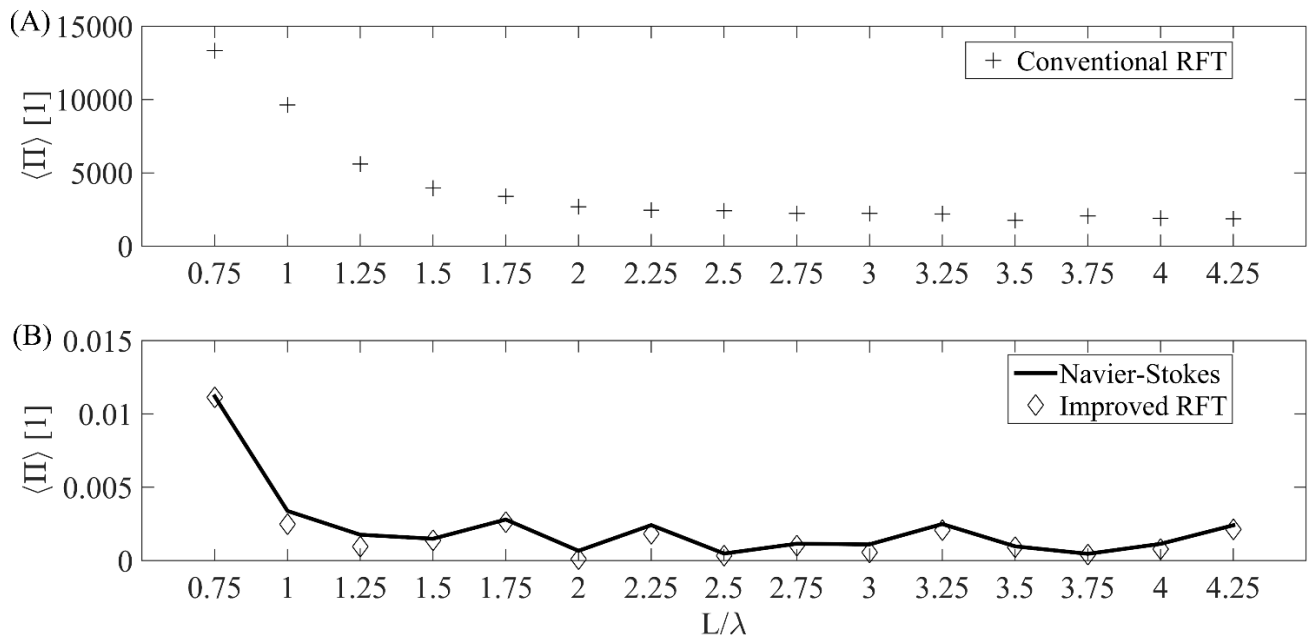

**Figure 8:** Time-averaged power comparisons for the beating tail: (**A**) results of the conventional RFT method; (**B**) results of the improved RFT analysis and CFD model governed by Navier-Stokes. All results are given with respect to varying wavelength. Axes are presented in the dimensionless form.

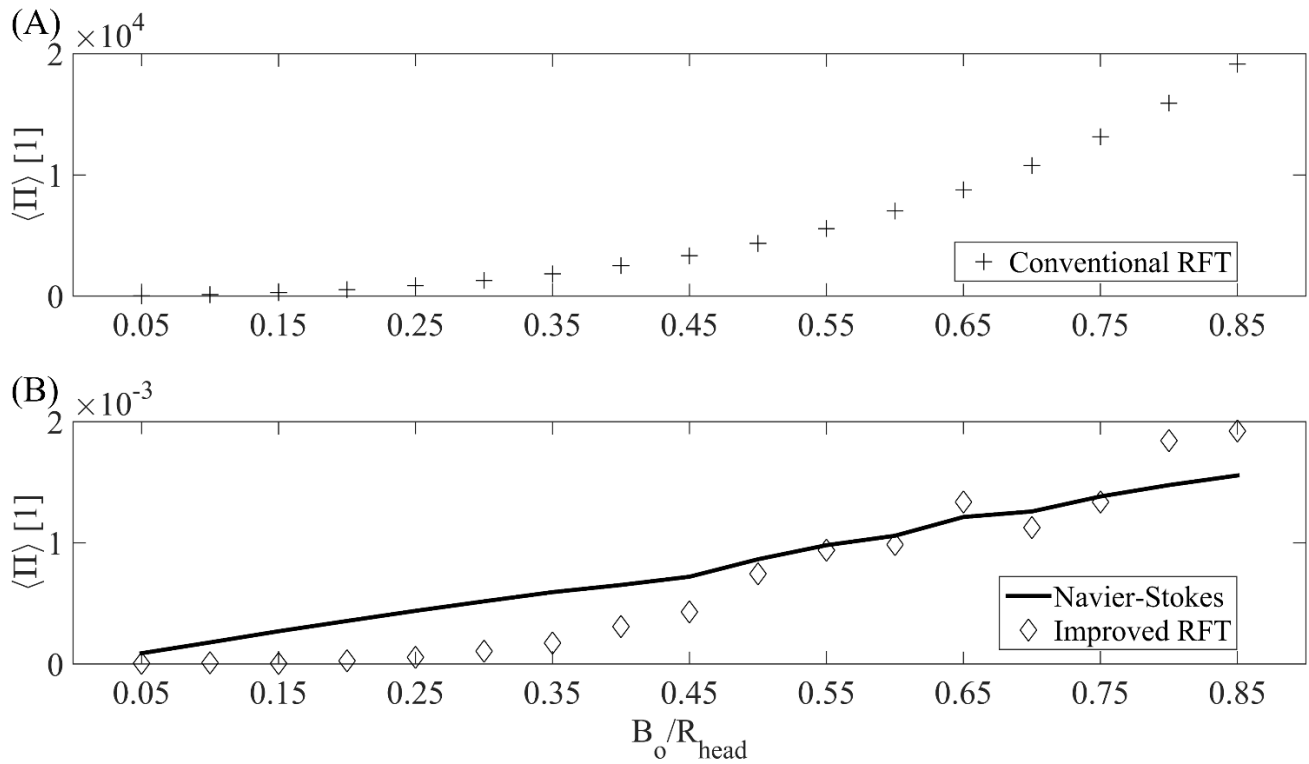

**Figure 9:** Time-averaged power comparisons for the beating tail: (**A**) results of the conventional RFT method; (**B**) results of the improved RFT analysis and CFD model governed by Navier-Stokes. All results are given with respect to varying wave amplitude. Axes are presented in the dimensionless form.

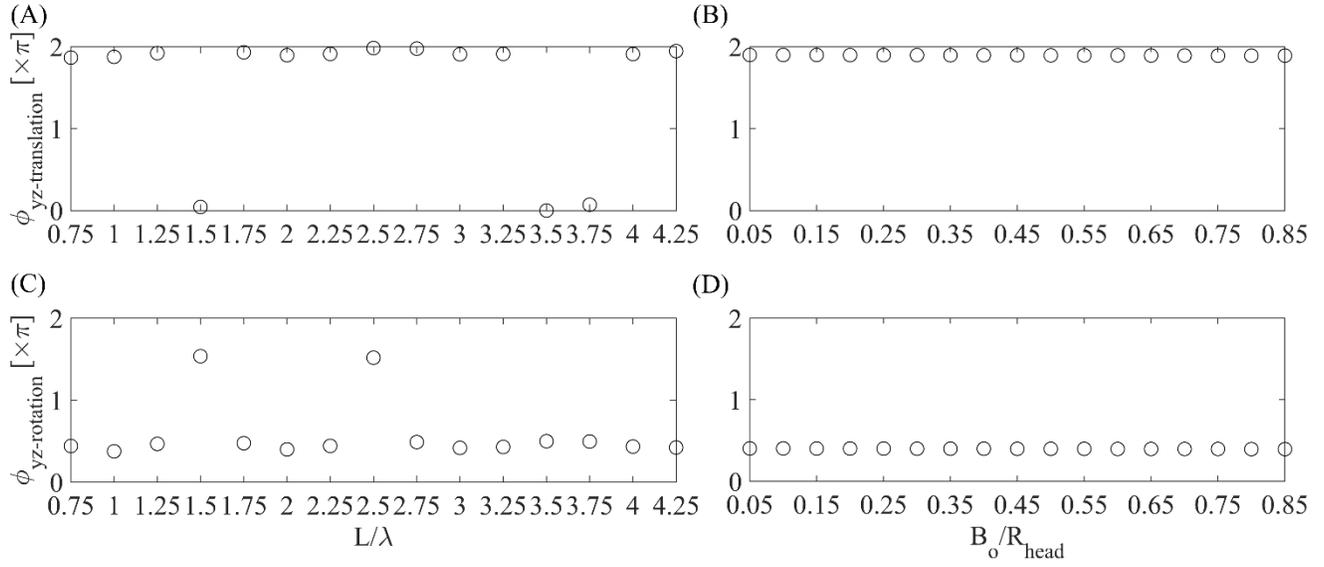

**Figure 10:** Time-averaged phase-correction coefficients for the resistance matrix of the head of the micro-swimmer: (**A**) the correction to the force associated with the lateral translation with respect to varying wavelength; (**B**) the correction to the force associated with the lateral translation with respect to varying wave amplitude; (**C**) the correction to the torque associated with the lateral rotation with respect to varying wavelength; (**D**) the correction to the torque associated with the lateral rotation with respect to varying wave amplitude. Axes are presented in the dimensionless form.

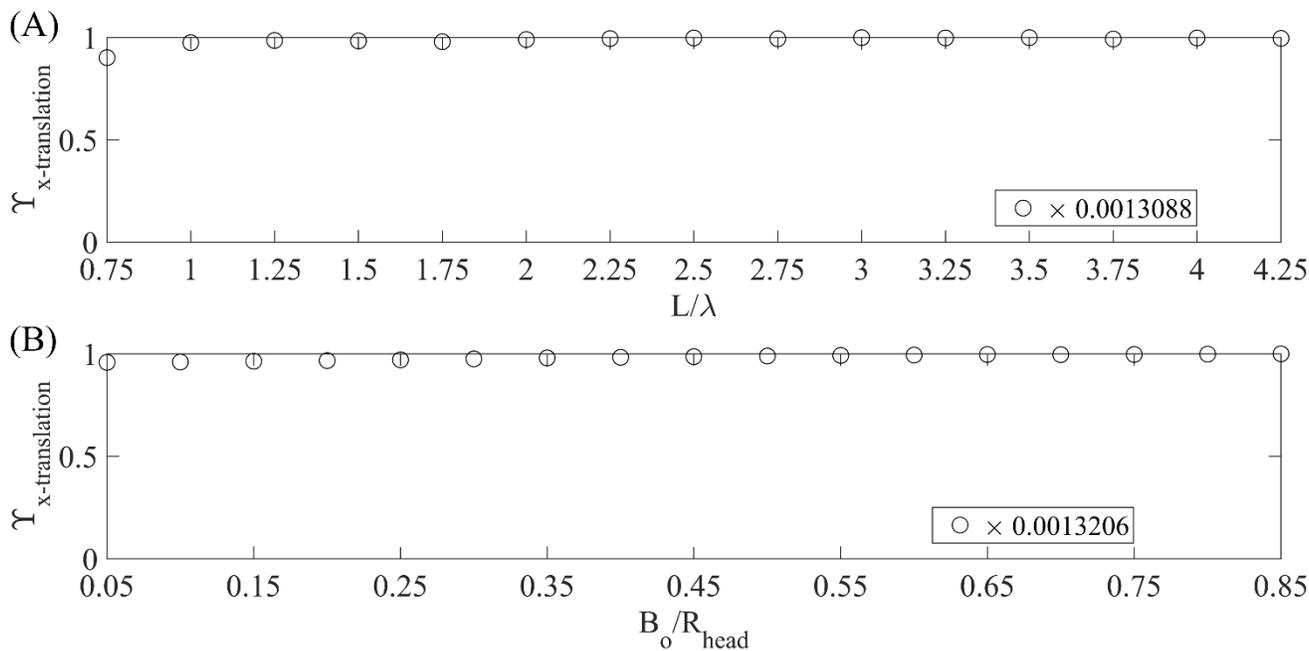

**Figure 11:** Time-averaged amplitude-correction coefficients for resistance matrix of the spherical head of the micro-swimmer; the correction for the force components along the direction of swimming: (**A**) the dependence on the wavelength; (**B**) dependence on the wave amplitude. Axes are presented in the dimensionless form.

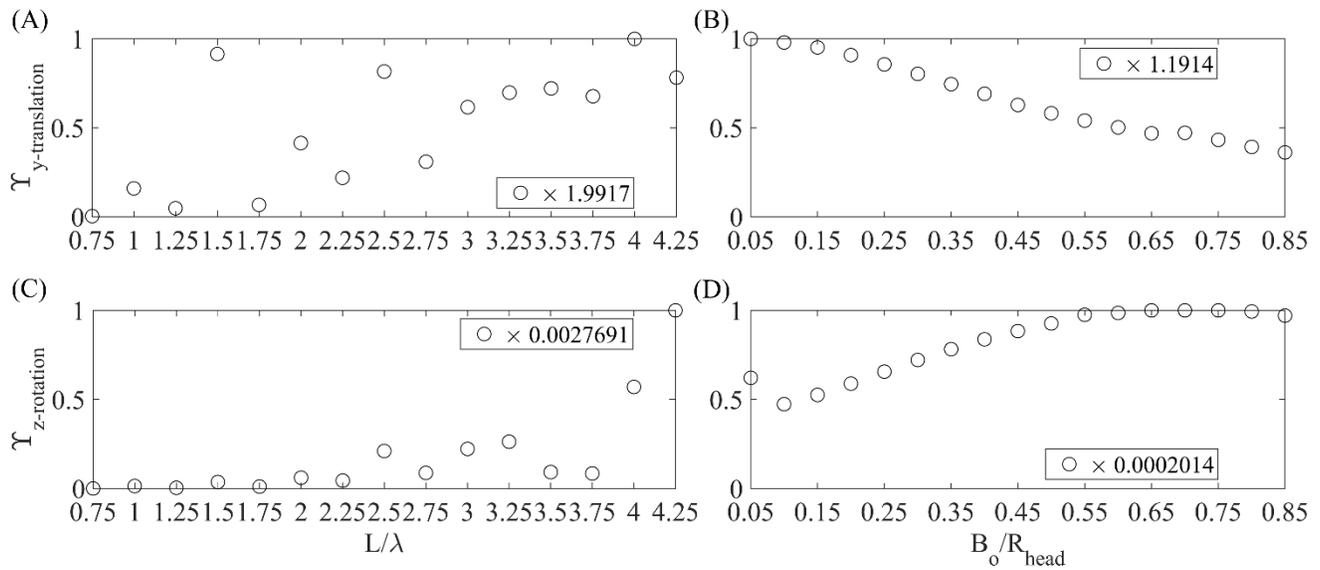

**Figure 12:** Time-averaged amplitude-correction coefficients for the resistance matrix of the head of the micro-swimmer: (**A**) the correction to the force associated with the lateral translation with respect to varying wavelength; (**B**) the correction to the force associated with the lateral translation with respect to varying wave amplitude; (**C**) the correction to the torque associated with the lateral rotation with respect to varying wavelength; (**D**) the correction to the torque associated with the lateral rotation with respect to varying wave amplitude. Axes are presented in the dimensionless form.

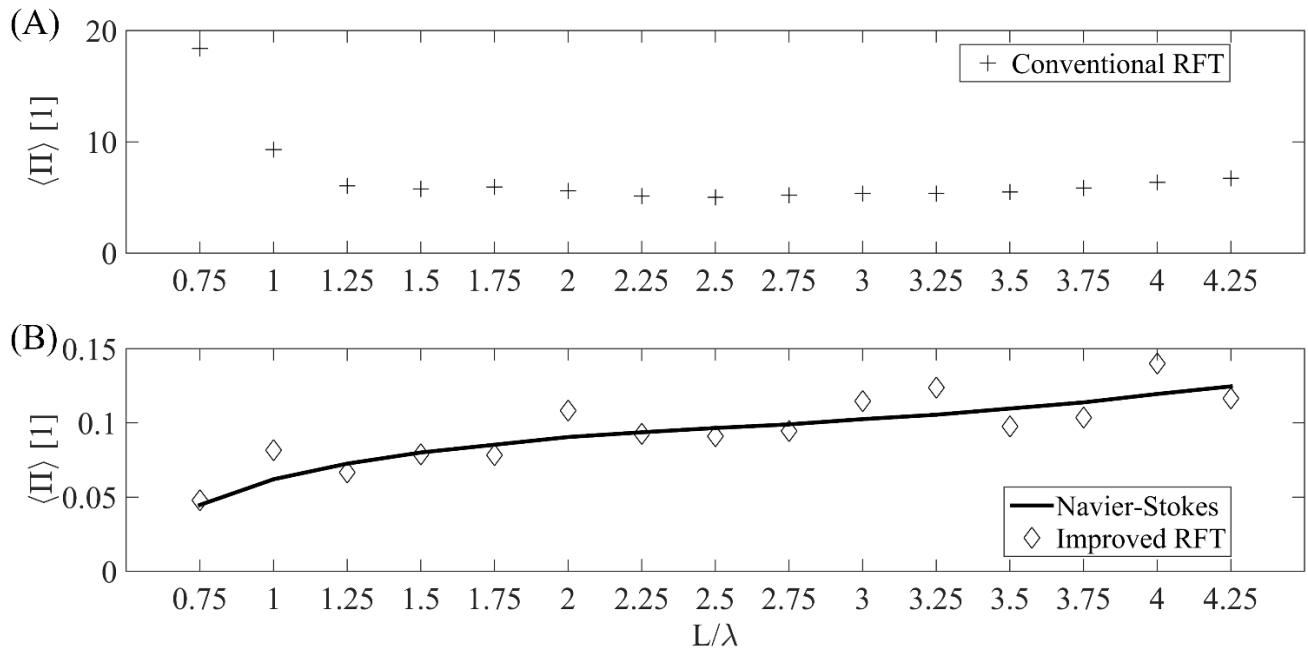

**Figure 13:** Time-averaged power comparisons for the spherical head of the micro-swimmer: (**A**) results of the conventional RFT method; (**B**) results of the improved RFT analysis and CFD model governed by Navier-Stokes. All results are given with respect to varying wavelength. Axes are presented in the dimensionless form.

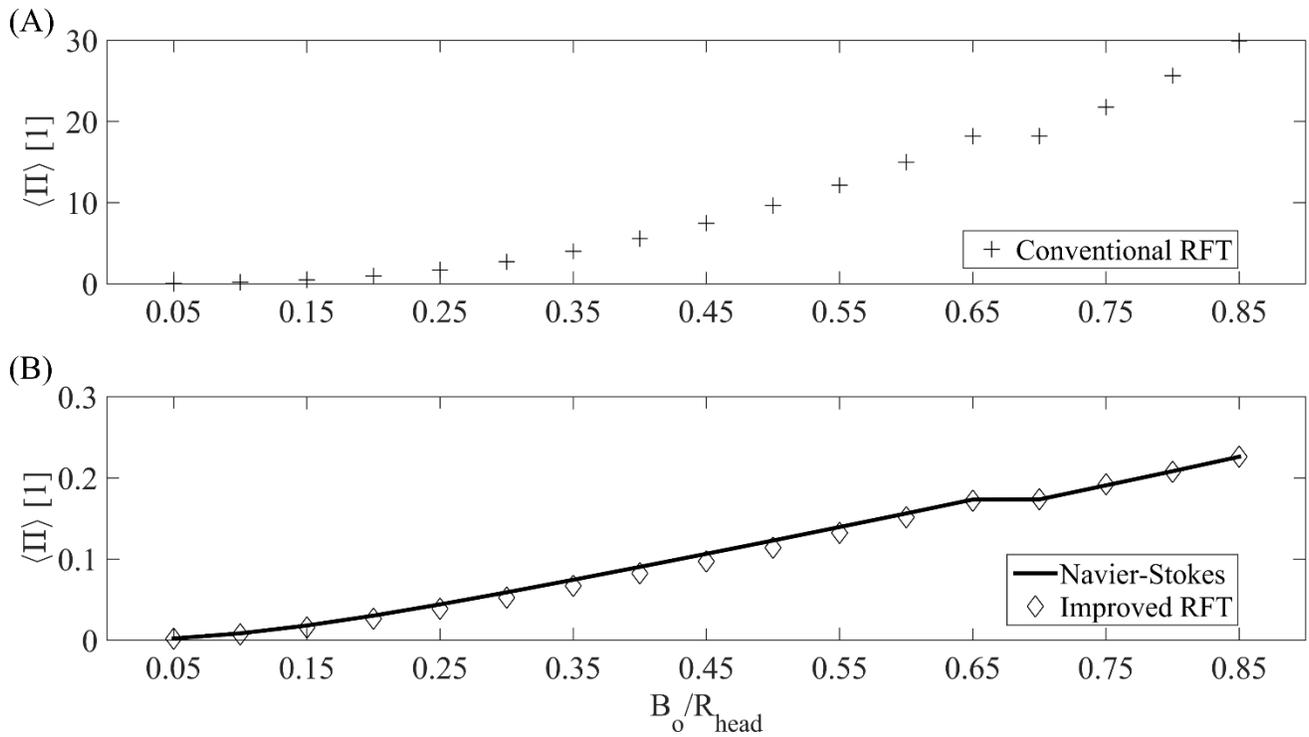

**Figure 14:** Time-averaged power comparisons for the spherical head of the micro-swimmer: (**A**) results of the conventional RFT method; (**B**) results of the improved RFT analysis and CFD model governed by Navier-Stokes. All results are given with respect to varying wave amplitude. Axes are presented in the dimensionless form.

## 11    Tables

Table 1: RMS Error in Power predictions of Conventional- and Improved-RFT approaches.

| | | Wavelength ($\lambda$) | | Wave Amplitude ($B_o$) | |
|---|---|---|---|---|---|
| | | Head | Tail | Head | Tail |
| Conventional RFT | Min | 4.9194 | $0.1764 \times 10^4$ | 0.0392 | $0.0024 \times 10^4$ |
| | Max | 18.3352 | $1.3338 \times 10^4$ | 29.7097 | $1.9140 \times 10^4$ |
| | Mean | 6.6716 | $0.3851 \times 10^4$ | 10.1077 | $0.5618 \times 10^4$ |
| | Standard Deviation | 3.3921 | $0.3331 \times 10^4$ | 9.5961 | $0.5978 \times 10^4$ |
| Improved RFT | Min | 0.0275 | $0.0740 \times 10^{-3}$ | 0.0013 | 0.0001 |
| | Max | 0.0826 | $0.9123 \times 10^{-3}$ | 0.1197 | 0.0011 |
| | Mean | 0.0588 | $0.3740 \times 10^{-3}$ | 0.0586 | 0.0006 |
| | Standard Deviation | 0.0151 | $0.2756 \times 10^{-3}$ | 0.0384 | 0.0003 |

**Errata for v.1:**

- Denominator of the RFT coefficient set is corrected.
- The misleading reference for the RFT coefficient set is corrected.
- Paranthesis added in $\Pi_{\text{CFD,tail,y}}(t)$ to avoid order of operation confusion.
- $z$-component of the tail amplitude correction terms is corrected.